\documentclass[prd,preprint,superscriptaddress,amsmath,amssymb]{revtex4}

 \pdfoutput=1

\usepackage{graphicx}
\usepackage{color,slashed}

\begin{document}
\title{\bf Scotogenic top-quark  FCNC decays }

\author{Chuan-Hung Chen}
\email{physchen@mail.ncku.edu.tw}
\affiliation{Department of Physics, National Cheng-Kung University, Tainan 70101, Taiwan}
\affiliation{ Physics Division, National Center for Theoretical Sciences, Taipei 10617, Taiwan.}

\author{Takaaki Nomura}
\email{nomura@scu.edu.cn}
\affiliation{College of Physics, Sichuan University, Chengdu 610065, China}

\date{\today}

\begin{abstract}
Flavor-changing neutral current (FCNC) top-quark decays are highly suppressed due to the Glashow-Iliopoulos-Maiani  mechanism in the standard model (SM). If $t\to q h, qV$ with $V=g,\gamma, Z$ are all induced via quantum loop levels, then we investigate the effect that can enhance the top-FCNC up to the  sensitivity designed at the high-luminosity (HL) LHC. Inspired by the mechanism of the scotogenic neutrino mass, we extend the SM by including  $Z_2$-odd colored fermions when a $Z_2$ discrete is imposed. The results show that by taking  $BR(t\to u g) \lesssim 0.61 \times 10^{-4}$ recently measured by ATLAS as an input, $t\to q \gamma$ can be indirectly bounded to be $BR(t\to q \gamma)\lesssim 3.2 \times 10^{-6}$, which is below the expected sensitivity at the HL LHC. After taking potential constraints from various experiments into account, the obtained branching ratios for the loop-induced $t\to q h$ and $t\to qZ$ decays can be $O(10^{-4})$, which falls within the sensitivity at the HL LHC. 

\end{abstract}
\maketitle

\section{Introduction}

Flavor-changing neutral currents (FCNCs) are suppressed at the tree level and can be induced via quantum corrections in the standard model (SM). However, loop-induced top-quark FCNCs are highly suppressed  in the SM due to  the Glashow-Iliopoulos-Maiani (GIM)  mechanism~\cite{Glashow:1970gm}.  As a result, the branching ratios (BRs) for the $t\to q (g, \gamma, Z, h)$ decays with $q=u, c$ in the SM are of the order  of $10^{-12}-10^{-17}$~\cite{AguilarSaavedra:2004wm,Abbas:2015cua,Balaji:2020qjg}, and the results are far below the LHC  sensitivities.

The expected sensitivities in the high-luminosity (HL)  LHC  with an integrated luminosity of 3ab$^{-1}$ at $\sqrt{s}=14$ TeV are expected to be $2.4-5.8\times 10^{-5}$ for $t\to q Z$~\cite{ATLAS:2019pcn}, $1.5\times 10^{-4}$ for $t\to q h$~\cite{ATLAS:2016qxw}, and $0.9-7\times 10^{-5}$ for $t\to q \gamma$~\cite{collaboration:2017gbu}. The event that the top-FCNC is found at the $10^{-5}$ level definitely indicates a new physics effect. Thus, the top-quark flavor-changing processes can  serve as good candidates for investigating new physics effects.  To explore the new physics effects in the rare top-quark decays, various extensions of the SM, which can reach the HL LHC sensitivity, were proposed in~\cite{Abraham:2000kx,Eilam:2001dh,AguilarSaavedra:2002kr,Gaitan:2017tka,Shen:2017oel,Chiang:2018oyd,Oyulmaz:2018irs,Chen:2018lze,Arroyo-Urena:2019qhl,Shi:2019epw,Liu:2020kxt,Hou:2020ciy,Bie:2020sro,Gutierrez:2020eby,Liu:2021crr,Cai:2022xha,Hernandez-Juarez:2022kjx,Badziak:2017wxn,Dey:2016cve}.

The  top-FCNC effects not only can be   detected via the new top-quark decay channels, but also  can be used to produce more top-quark events, such as $g q \to t$. Using the single-top production, the upper limits on the BR using 139 fb$^{-1}$ at $\sqrt{s}=13$ TeV in ATLAS are shown as $BR(t\to u g) < 0.61 \times 10^{-4}$ and $BR(t\to c g) < 3.7 \times 10^{-4}$~\cite{ATLAS:2021amo}. 


 If all mentioned  top-FCNC processes are enhanced via quantum loop effects, the intermediate states in the loops  may carry the quantum number that the SM particles do not possess. 
A known example is the radiative neutrino mass in  a scotogenic model~\cite{Ma:2006km}, where  the  mediated particles in the loop are   $Z_2$-parity odd, whereas the SM particles are $Z_2$-parity even.  Interestingly, in addition to the explanation of the neutrino mass, the predicting  dark matter (DM) candidate can fit the DM relic density that was observed by the Planck collaboration~\cite{Planck:2018vyg}. 

 If we have $Z_2$-odd colorless fermions, it is reasonable that there also exist $Z_2$-odd colored fermions, which are similar to the quarks in the SM.
 Hereafter, we call these particles as  $Z_2$-odd quarks.  Based on this assumption, when  we make a minimal extension of the Ma-model proposed in~\cite{Ma:2006km}, we investigate  if the top-FCNC processes can be enhanced up to  the sensitivities of the HL LHC. As the minimal requirement,  no new local gauge symmetry is considered, and the number of including $Z_2$-odd quarks is as less as possible.  Since a new local gauge symmetry introduces a new gauge coupling and gauge boson(s), which are not directly related to the top-FCNC processes, thus, we only focus on the SM gauge symmetry. 

Because the $t\to q \gamma (h)$  decay involves the structure of a dipole (scalar) current, the chirality in the initial and final quarks has to be different. In addition, the left-handed and right-handed top quarks are $SU(2)$ doublet and singlet in the SM, respectively. Therefore, to avoid the chirality suppression of $m^2_t/\Lambda^2$, where $\Lambda$ is the mass scale of the new heavy particle,  the candidate of new colored fermions should be chosen in such a way that the chirality suppression can be overcome.  Without introducing extra new $Z_2$-odd scalar field with the exception of the inert Higgs doublet, we find that the possibly minimal  representations for the colored fermions in $SU(2)_L\times U(1)_Y$ gauge symmetry are the vector-like doublet and singlet. Vector-like particles  are used because their gauge anomaly can be evaded.  

Three scalar bosons exist in an inert Higgs doublet, namely, inert charged Higgs, scalar, and pesudoscalar, where the lightest neutral inert scalar can be the DM candidate.   Due to the suppression factor of $m^2_t/\Lambda^2$, which arises from the chirality flip of top-quark, the contributions from the neutral inert scalars to $t\to qh$ are small; therefore, the $t\to q h$ processes are dominated by the inert charged-Higgs. Although $t\to q \gamma$ can avoid the chirality suppression,  since the loop-induced effect is associated with the mixing between $Z_2$-odd doublet and singlet quark, which is of order of $v/\Lambda$ with $v$ being the vacuum expectation value of Higgs field, $t\to q \gamma$ dominated by the inert charged Higggs in the model cannot be enhanced up to the sensitivity of HL LHC. For the $t\to q Z$ processes, both neutral and charged scalars can have significant contributions. 


The remainder of this paper is organized as follows: We introduce the model and derive the relevant Yukawa and gauge couplings in Section~\ref{sec:model}. Based on the obtained couplings, we formulate the decay amplitudes and BRs for the studied  top decays in Section~\ref{sec:amplitudes}. Then, we discuss various possible constraints, which include radiative $B$-meson decay, oblique parameters, Higgs production, and DM detections, in Section~\ref{sec:constraints}. We analyze and discuss the numerical results in detail in Section~\ref{sec:num}. Finally, we summarize the study in Section~\ref{sec:summary}.

\section{ Model and the new couplings }\label{sec:model}

 To enhance the top-quark FCNC processes through radiative corrections without introducing a new gauge symmetry in a scotogenic mechanism, we impose a $Z_2$ discrete symmetry in the model and  extend the SM, including a new Higgs doublet $(\Phi_I)$, three singlet Majorana fermions $N_{R i}$ ($i=$1-3) and vector-like quark doublet ($Q_4$) and  singlet ($B'$), where the introduced particles are  $Z_2$-odd and the SM particles are $Z_2$-even under the $Z_2$ transformation.   The representations and charge assignments of new particles in $SU(2)_L \times U(1)_Y \times Z_2$ are shown in Table.~\ref{tab:rep}. 
The $Z_2$ odd Majorana fermions $N_R$s are necessary to generate neutrino mass by scotogenic mechanism. In this work we do not discuss neutrino mass generation since it is exactly the same as original Ma-model~\cite{Ma:2006km}.  Due to the unbroken $Z_2$ symmetry,  the neutral component of the inert Higgs doublet and the lightest Majorana fermion can be the DM
 candidate~\cite{Barbieri:2006dq}.  Since the inert Higgs doublet is directly related to the top-FCNC process, in this study, we take the lightest neutral component of $\Phi_I$ as the DM candidate.  
 We derive the relevant couplings for top-quark FCNC in the following discussions.

 \begin{table}[htp]
\caption{Representations and charge assignments for new particles.}
\begin{center}
\begin{tabular}{c|cccc} \hline
  ~~& ~~~~ $ SU(3)_C$~~~~ & ~~~~ $ SU(2)_L$~~~~ & ~~~~$U(1)_Y$~~~~  & ~~ $Z_2$ \\ \hline
 $Q_4$ ~~& $3$ & $2$  & $1/3$  & $-1$ \\ \hline 
 $B'$ ~~& $3$ & $1$  & $-2/3$  & $-1$ \\ \hline 
 $\Phi_I$ ~~& $1$ & $2$ & $1/2$ & $-1$ \\ \hline
 $N_{R i}$ ~~& $1$ & $1$ & $0$ & $-1$ \\ \hline
    
\end{tabular}
\end{center}
\label{tab:rep}
\end{table}%

\subsection{Yukawa couplings and scalar potential} 

 Based on the charge assignments shown in Table~\ref{tab:rep}, the new Yukawa interactions can be written as
 \begin{align}
 -{\cal L}_Y & = y_{B'}  \overline{Q_{4L}} H B'_R + \tilde{y}_{B'} \overline{Q_{4R}}  H B'_L  +  \overline{Q_{4L}} {\bf Y}^d_{I} \Phi_I d_R  +  \overline{Q_{4L}} {\bf Y}^u_{I}  \widetilde{\Phi_I}  u_R \nonumber \\ 
&    + \overline{Q_L} {\bf Y}^{B'}_{I} \Phi_I B'_R + m_{4Q} \overline{Q_{4L}} Q_{4R} + m_{B'} \overline{B'_L} B'_R + H.c.\,,  \label{eq:newYu}
 \end{align}
where the flavor indices are suppressed; $\widetilde{\Phi_I}= i \tau_2 \Phi^*_I$ and $\tau_2$ is the Pauli matrix, and $H$ is the SM Higgs doublet.  In this work, we considered that  the new Yukawa couplings are real. Since Eq.~(\ref{eq:newYu}) does not involve the SM quark mass diagonalization, the up- and down-type quarks can be taken as the physical states, and the flavor mixing effects are absorbed into the Yukawa couplings. 
The doublet components of $Q_{4}$, $\Phi_I$, and $H$ are taken as
 \begin{equation}
 Q_{4}  = 
\left(
\begin{array}{c}
 T     \\
  B      \\    
\end{array}
\right)\,, ~~ \Phi_{I} = \left(
\begin{array}{c}
 H^+_I   \\
  \frac{1}{\sqrt{2}} (H_I + i A_I)      \\    
\end{array}
\right)\,, ~~ H = \left(
\begin{array}{c}
 G^+   \\
  \frac{1}{\sqrt{2}} (v+h + i G^0)      \\    
\end{array}
\right)\,.
 \end{equation}
Thus, the first two terms in Eq.~(\ref{eq:newYu}) lead to the mixture between $B$ and $B'$, and the associated mass matrix is expressed as
 \begin{equation}
 (\bar B', \bar B)_L  \left(
\begin{array}{cc}
m_{B'} &   \frac{\tilde{y}_{B'} v}{\sqrt{2}}   \\
  \frac{y_{B'} v}{\sqrt{2}}   &  m_{4Q}    \\    
\end{array}
\right)  \left(
\begin{array}{c}
 B'     \\
  B      \\    
\end{array}
\right)_R\,.  \label{eq:mass_B}
 \end{equation}
 In general, we need bi-unitary transformation to diagonalize the mass matrix in Eq.~(\ref{eq:mass_B}).  In order to simplify the analysis,  we take $\tilde{y}_{B'}=y_{B'}$ and the $2\times 2$ real symmetric matrix 
  can be diagonalized by an $SO(2)$ transformation, where the eigenvalues and eigenstates can be obtained as 
 \begin{align}
 m_{B_{1(2)}} &  = \frac{m_{4Q} + m_{B'}}{2} \mp \frac{1}{2} \sqrt{(m_{4Q} -m_{B'})^2 + 2 y^2_{B'} v^2}\,, \nonumber \\
  \left(
\begin{array}{c}
 B_1     \\
  B_2      \\    
\end{array}
\right) &=  \left(
\begin{array}{cc}
\cos\theta &   -\sin\theta   \\
  \sin\theta    & \cos\theta    \\    
\end{array}
\right)  \left(
\begin{array}{c}
 B'     \\
  B      \\    
\end{array}
\right)\,, \nonumber \\
\tan2\theta & = \frac{\sqrt{2} y_{B'} v}{m_{4Q} - m_{B'}}\,, \label{eq:massBBp}
 \end{align}
 where $\theta$ is the mixing angle of $B$ and $B'$. 
 We have taken $B_1$ as the lightest $Z_2$-odd quark. 
The resulting Yukawa couplings  are then written as
 \begin{align}
  {\cal L}_{Y} & \supset   \bar B_1 \left (- s_\theta {\bf Y}^u_I P_R - c_\theta {\bf Y}^{B'}_I P_L \right) u H^-_I  +  \bar B_2 \left ( c_\theta {\bf Y}^u_I P_R - s_\theta {\bf Y}^{B'}_I P_L \right) u H^-_I \nonumber \\
  & - \frac{1}{\sqrt{2}} \bar T {\bf Y}^u_I P_R u ( H_I - i A_I) -  \frac{1}{\sqrt{2}} (c_\theta \bar B_1 + s_\theta \bar B_2)  {\bf Y}^{B'}_I P_L d ( H_I - i A_I)  \nonumber \\
  & - \bar T \frac{{\bf Y}^d_I}{\sqrt{2} } P_R\, d\, H^+_I - \left(-s_\theta \bar B_1 + c_\theta \bar B_2 \right) \frac{{\bf Y}^d_I}{\sqrt{2} } P_R \, d \left( H_I + i A_I\right)+ H.c.\,, \label{eq:Yukawa}
 \end{align}
with $c_\theta(s_\theta)=\cos\theta(\sin\theta)$.  Using the parametrization, the Higgs couplings to $B_1$ and $B_2$ can be written as 
 \begin{align}
 {\cal L}_{hBB}& = -\frac{ 1 }{\sqrt{2}} h (\bar B_1, \bar B_2) {\bf y}^h \left( \begin{array}{c}
 B_1    \\
  B_2      \\    
\end{array}
\right)  \nonumber \\
& = -
 \frac{ y_{B'} }{\sqrt{2}} h (\bar B_1, \bar B_2)   \left(
\begin{array}{cc}
 - s_{2\theta} &   c_{2\theta}   \\
 c_{2\theta}   &   s_{2 \theta}    \\    
\end{array}
\right)  \left(
\begin{array}{c}
 B_1    \\
  B_2      \\    
\end{array}
\right)\,, \label{eq:hBB}
 \end{align}
with $c_{2\theta} (s_{2\theta}) =\cos2\theta (\sin2\theta) $. The $hB_i B_i$ coupling, which will lead to the $t\to q h$ decay at the loop level, can be induced from the mixing terms of $B_1$ and $B_2$. We note that there is no $hTT$ coupling  at the tree level in the model. 

The masses of the inert scalars  and  their couplings to the Higgs are determined by the scalar potential, which can be written as~\cite{Ma:2006km,Barbieri:2006dq}
 \begin{align}
 V(H, \Phi_I) & = \mu^2_1 H^\dag H + \mu^2_2 \Phi^\dag_I \Phi_I + \lambda_1 (H^\dag H )^2 + \lambda_2 (\Phi^\dag_I  \Phi_I )^2 \nonumber \\
 & + \lambda_3 (H^+ H) (\Phi^\dag_I \Phi_I) + \lambda_4 (H^+ \Phi_I) (\Phi^\dag_I H) + \left[ \frac{1}{2} \lambda_5 (H^\dag \Phi_I)^2 + H.c.\right]\,.
 \end{align}
 We can obtain the mass squares of $H^\pm_I$, $H_I$, and $A_I$ as
  \begin{align}
      m^2_{H_I} & = \mu^2_2 + \frac{\lambda_L v^2}{2} \,,~
     m^2_{A_I}  = \mu^2_2 + \frac{\lambda_A v^2}{2}\,,  ~
      m^2_{H^\pm_I}  = \mu^2_2 + \frac{\lambda_3 v^2}{2} \,, \label{eq:massHI}
  \end{align}
with $\lambda_{L(A)}=\lambda_3 + \lambda_4 \pm \lambda_5$. The mass difference between $H_I$ and $A_I$ depends on  the $\lambda_5$ parameter. The  Higgs couplings to $(H^\pm_I, H_I, A_I)$ can be determined as
 \begin{align}
 hH^+_I H^-_I : \lambda_3\, v \,,  ~ ~ h H_I H_I :  \lambda_L  v \,,~ ~ h A_I A_I : \lambda_A v. \label{eq:hHIHI}
 \end{align}

 In addition to the minimal conditions, i.e., $\partial V/\partial h(H_I) =0$, the vacuum stability is controlled by the copositivity criteria in the dimension-4 terms of scalar potential, and the stable conditions are yielded as~\cite{Barbieri:2006dq,Klimenko:1984qx,Kannike:2012pe}
 \begin{align}
 \lambda_{1,2} \geq 0\,,  ~\lambda_3 + 2 \sqrt{\lambda_1 \lambda_2} \geq 0\,, ~ \lambda_3 + \lambda_4 - |\lambda_5| +  2 \sqrt{\lambda_1 \lambda_2} \geq 0\,. \label{eq:BFB}
 \end{align}
 If we take the  mass ordering to be $m_{H^\pm_I} > m_{A_I} > m_{H_I}$, then $\lambda_{4,5} < 0$, which can be  expressed as
  \begin{equation}
  |\lambda_4| = \frac{1}{v^2} \left( 2 m^2_{H^\pm_I} - m^2_{H_I} -m^2_{A_I} \right)\,, ~ |\lambda_5| = \frac{m^2_{A_I} -m^2_{H_I}}{v^2}\,.
  \end{equation}
%
%
To obtain the potential bounded from below, where the conditions in  Eq.~(\ref{eq:BFB}) are satisfied, we  require $\lambda_3 >0$. Thus,  some cancellation occurs in the  $h S_I S_I$ ($S_I=H_I, A_I)$ coupling.  In addition, in Eq.~(\ref{eq:Yukawa}),  $S_I  T u$ couplings are only associated with  the right-handed up-type SM quarks.  As a result,   the  loop-induced $t\to q (h, \gamma)$ processes through the $S_I  T u$ couplings are suppressed by $m^2_t/m^2_T$, and they are negligible if $m_T\sim O(1)$ TeV. Hence, when the heavy quark and inert Higgs masses are fixed, the main parameters that affect the rare top decays in the model are  the Yukawa couplings ${\bf Y}^{B'}_{I}$ and ${\bf Y}^u_I$, the mixing angle $\theta$,  and the $\lambda_3$ parameter that is directly related to the $hH^-_I H^+_I$ coupling.

 \subsection {Gauge couplings to new fermions and scalars}

To study the $t\to q V$ ($V=\gamma, Z$) processes, we also need the gauge couplings of  the photon and $Z$-boson to the inert Higgses and to the $Z_2$-odd quarks,  where the gauge couplings   from the kinetic terms of $H_I$, $Q_4$, and $B'$ are written as
 \begin{align}
 {\cal L}_{\rm gauge} \supset (D_\mu \Phi_I)^\dag D^\mu \Phi_I +   \overline {Q_{4}}  i \slashed{D}  Q_4 + \overline {B'} i \slashed{D'}  B'  \,.
 \end{align}
 The covariant derivatives for the  doublet and singlet fields are taken as
 \begin{align}
 D^\mu &= \partial^\mu + i g \vec{T} \cdot \vec{W}^\mu + i g' \frac{Y}{2} B_\mu \,, \nonumber \\
 D'^\mu &= \partial^\mu + i g' \frac{Y}{2} B_\mu\,,
 \end{align}
 where $W^{i\mu}$ and  $B^\mu$ are the gauge fields of $SU(2)_L$ and $U(1)_Y$, respectively.  Using the standard notations for the photon and $Z$-boson, which are defined as
  \begin{align}
  A_\mu & = c_W B_\mu + s_W W^3_\mu\,, \nonumber \\
  Z_\mu & = - s_W B_\mu +c_W W^3_\mu \,.
  \end{align}
 The neutral gauge couplings to the inert Higgses are obtained as
  \begin{align}
  {\cal L}_{\rm VS_I S_I} \supset i \left( e A^\mu + \frac{g c_{2W} } { 2 c_W}  Z^\mu \right)   ( H^+_I \partial_\mu H^-_I   - H^-_I \partial_\mu H^+_I )  +   \frac{g }{2c_W} Z^\mu ( A_I \partial H_I - H_I \partial_\mu A_I ) \,, \label{eq:VSS}
  \end{align}
 where  $c_{W}(s_W)=\cos\theta_W(\sin\theta_W)$, and $\theta_W$ is the Weinberg's angle.
  Since the quark $T$ doesn't mix with the SM up-type quarks, its gauge couplings to the photon and the $Z$ boson can be simply written as
   \begin{equation}
  -  e Q_t \bar T \gamma_\mu T A^\mu   - \frac{g}{c_W} \left( \frac{1}{2} -Q_t s^2_W \right) \bar T \gamma_\mu  T Z^\mu \,.  \label{eq:TT}
   \end{equation}
  Although the photon couplings to $B_1$ and $B_2$  are flavor-conserved, since  the $B$ and $B'$ quarks belong to  different $SU(2)_L$ representations and are mixed via the Yukawa couplings, their $Z$ gauge couplings allow flavor-changing and are expressed as
  \begin{align}
  {\cal L}_{VBB} &= - e Q_b \sum_i \bar B_i \gamma_\mu B_i A^\mu - \frac{g}{c_W} (\bar B_1 , \bar B_2)  \gamma_\mu {\bf C}^Z  \left(
\begin{array}{c}
 B_1    \\
  B_2      \\    
\end{array}
\right) Z^\mu \,, \label{eq:VBB}
  \end{align}
 with
  \begin{equation}
  {\bf C}^Z = \left(
\begin{array}{cc}
 - s^2_\theta/2 - Q_b s^2_W &  s_\theta c_\theta    \\
 s_\theta c_\theta  &   - c^2_\theta/2 - Q_b s^2_W  \\    
\end{array}
\right)\,.
  \end{equation}
  Because the charged gauge boson $W^\pm$ only couples to the doublet quarks, the $W^\pm$ coupling can be found as
   \begin{equation}
   {\cal L}_{WTB} = - \frac{g}{\sqrt{2}} \bar T \gamma_\mu \left( -s_\theta B_1 + c_\theta B_2 \right) W^{+\mu} + H.c.
   \end{equation}
  
\section{ Loop-induced decay amplitudes for $t\to q (h, V)$} \label{sec:amplitudes}

In this section,  we derive the decay amplitudes for the $t\to q (h, V)$ decays. Because the calculations for $t\to q g$ are similar to  those for $t\to q \gamma$,  and their BRs can be approximately related by $BR(t\to q g) \sim C_F (\alpha_s/\alpha) BR(t\to q \gamma)$ with $C_F=4/3$ in the model, where $\alpha=e^2/4\pi$ and $\alpha_s=g^2_s/4\pi$. In this study,  we only focus on the $t\to q \gamma$ analysis.

\subsection{ $t\to q (h, \gamma )$ decays}

\begin{figure}[phtb]
\begin{center}
\includegraphics[scale=0.65]{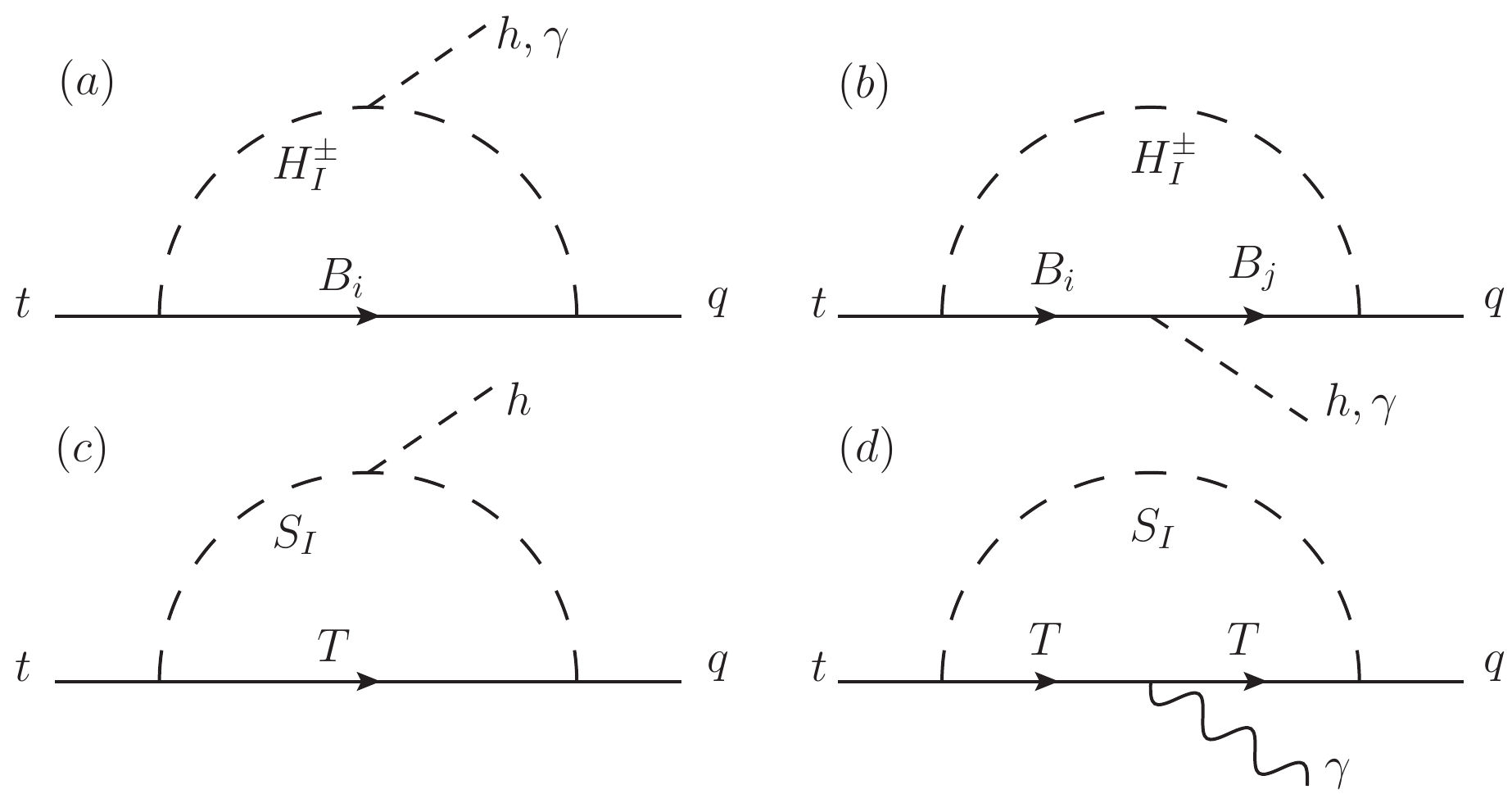}
 \caption{ Feynman diagrams used to induce the $t\to q (h, \gamma)$ processes, where the emitted dashed lines can be the Higgs or photon. }
\label{fig:ttoqh}
\end{center}
\end{figure}

In the model, the $t\to q ( h, \gamma)$ processes can be induced from the loops with the mediation of $H^\pm_I$ and $S_I$. The Feynman diagrams are sketched in Fig.~\ref{fig:ttoqh},  where the emitted dashed  lines can be the Higgs or the photon. Since there is  no $hTT$ coupling, the $t\to q h$ decay cannot be produced from Fig.~\ref{fig:ttoqh}(d). In Fig.~\ref{fig:ttoqh}(b), the flavor-changing between $B_1$ and $B_2$ only occurs in the Higgs coupling. 

The  effective interactions for $t\to q (h, \gamma)$ can be written as
 \begin{equation}
 {\cal L}_{t\to q (h, \gamma)}  = - C^h_{L} \bar q P_L t\,  h - C^h_R \bar q P_R t\, h+ \frac{ e B^\gamma_L}{m_t}  \bar q i \sigma_{\mu \nu} \epsilon^{\mu*}_\gamma k^\nu  P_L  t +  \frac{e B^\gamma_R}{m_t}  \bar q i \sigma_{\mu \nu} \epsilon^{\mu*}_\gamma k^\nu  P_R t \,.
 \end{equation}
 Following the Feynman diagrams shown in Fig.~\ref{fig:ttoqh} and the couplings obtained  in Eq.~(\ref{eq:Yukawa}) and Eqs.~(\ref{eq:VSS}-\ref{eq:VBB}), the effective Wilson coefficients with $\chi=L,R$ can be written as
 \begin{align}
 C^h_{\chi} & = \sum_i \left( C^{h a}_{i \chi} + \sum_j C^{h b}_{ij \chi} \right) + C^{h c}_\chi\,, \nonumber \\ 
 B^\gamma_\chi & = \sum_{i} \left( B^{\gamma a}_{i\chi}  + B^{\gamma b}_{i\chi} \right)+  B^{\gamma d}_{\chi}\,.
 \end{align}
The contributions to $t\to q h$  from each diagram can be formulated as
 \begin{align}
 C^{ha}_{iL} & = \frac{\lambda_3 v}{ (4\pi)^2 m_t } C^q_{iR} \int^1_0 dx_1 \int^{x_1}_0 dx_2 \frac{C^t_{iR} r_{it} x_2 + \sqrt{r_{it} }C^t_{iL} }{D^a_h(r_{i H^\pm_I}, r_{it}, r_{ih} )} \,, \nonumber \\
   C^{ha}_{iR} & = \frac{\lambda_3 v}{ (4\pi)^2 m_t } C^q_{iL} \int^1_0 dx_1 \int^{x_1}_0 dx_2 \frac{C^t_{iL} r_{it} x_2 + C^t_{iR}  \sqrt{r_{it} }}{D^a_h(r_{i H^\pm_I}, r_{it}, r_{ih} )} \,, \nonumber \\
 C^{hb}_{ijL} & = - \frac{y^h_{ij} }{(4\pi)^2}   \int^1_0 dx_1 \int^{x_1}_0 dx_2  \frac{C^q_{jR}}{D^b_h(r_{j H^\pm_I}, r_{jt}, r_{ji}, r_{jh})} \left[ C^t_{iR}(r_{jt} x_2 (1-x_1) \right. \nonumber \\
 & \left. + r_{jh} (x_1 -x_2) -r_{ji}) + C^t_{iL} \sqrt{r_{jt} } (\sqrt{r_{jt}} x_2 - (1-x_2) ) \right]\,, \nonumber \\
 C^{hb}_{ijR} & = - \frac{y^h_{ij} }{(4\pi)^2}   \int^1_0 dx_1 \int^{x_1}_0 dx_2  \frac{C^q_{jL}}{D^b_h(r_{j H^\pm_I}, r_{jt}, r_{ji}, r_{jh})} \left[ C^t_{iL}(r_{jt} x_2 (1-x_1) \right. \nonumber \\
 & \left. + r_{jh} (x_1 -x_2) -r_{ji}) + C^t_{iR} \sqrt{r_{jt} } (\sqrt{r_{jt}} x_2 - (1-x_2) ) \right]\,, \nonumber \\
 C^{hc}_{L} & = \frac{Y^u_{Iq} Y^u_{I3} y_t }{ (4\pi)^2} \frac{v }{m_t}    \int^1_0 dx_1 \int^{x_1}_0 dx_2\,  x_2 \left[ \frac{\lambda_3 + \lambda_4 + 2 \lambda_5 }{D^a_h( y_{ H_I}, y_{t}, y_{h} )}  + \frac{\lambda_3 + \lambda_4 +-2 \lambda_5 }{D^a_h( y_{A_I}, y_{t}, y_{h} )} \right]\,, \label{eq:Ch}
 \end{align}
 where $r_{if} = m^2_{f}/m^2_{B_i}$, $y_f=m^2_f/m^2_T$; the Yukawa couplings $C^f_{i\chi}$ are
 \begin{align}
 C^f_{1R} & = -s_\theta Y^u_{If}\,,~ C^f_{1L} = -c_\theta Y^{B'}_{If} \,,  \nonumber \\
 C^f_{2R} & = c_\theta Y^u_{If} \,, ~C^f_{2L} = -s_\theta  Y^{B'}_{If}\,,
 \end{align}
 and the denominators are defined by
 \begin{align}
 D^{a}_h (x,y,z)&= 1 - (1- x ) x_1  - y (1- x_1) x_2 - z (x_1 -x_2) x_2 \,, \nonumber \\
  D^{b}_h (w,x,y,z)&= x_1 + w (1- x_1) - (x-y+1) x_2 + z x^2_2 + (x-z) x_1 x_2\,. 
 \end{align}
  The results for the  $t\to q \gamma$ decay from each Feynman diagram are obtained as
\begin{align}
  B^{\gamma a}_{iL} & = -\frac{1}{(4\pi)^2} \int^1_0 dx_1 \int^{x_1}_{0} dx_2 \frac{1- x_1}{ D^{a}_\gamma (r_{i H^\pm_I},r_{it}) } C^q_{iR}  \left[ 
      C^t_{iR} r_{it} x_1 +  C^t_{iL} \sqrt{r_{it}}   \right ]\,, \nonumber \\
 B^{\gamma a}_{iR} & = - \frac{1}{(4\pi)^2}\int^1_0 dx_1 \int^{x_1}_{0} dx_2 \frac{1- x_1}{ D^{a}_\gamma (r_{i H^\pm_I}, r_{it}) } C^q_{iL}  \left[ 
      C^t_{iL} r_{it} x_1 + C^t_{iR}  \sqrt{r_{it}}  \right ]\,, \nonumber \\
    B^{\gamma b}_{iL} & =\frac{ Q_b}{(4\pi)^2}  \int^1_0 dx_1 \int^{x_1}_{0} dx_2 \frac{1}{ D^{b}_\gamma ( r_{i H^\pm_I}, r_{it})} C^q_{iR}  \left[ 
      C^t_{iR} r_{it} (1-x_1) x_2 + C^t_{iL} \sqrt{r_{it}}  x_1  \right ]\,, \nonumber \\
 B^{\gamma b}_{iR} & = \frac{ Q_b}{(4\pi)^2}  \int^1_0 dx_1 \int^{x_1}_{0} dx_2 \frac{1}{  D^{b}_\gamma ( r_{i H^\pm_I}, r_{it} ) } C^q_{iL}  \left[ 
      C^t_{iL} r_{it} (1-x_1) x_2  + C^t_{iR} \sqrt{r_{it}}  x_1  \right ]\,, \nonumber \\
   B^{\gamma d}_{L} & = \frac{Q_t Y^u_{Iq} Y^u_{I3}}{2 (4\pi)^2}     y_t \int^1_0 dx_1 \int^{x_1}_0 dx_2 \, x_2 (1-x_1) \left[ \frac{1 }{ D^{b}_\gamma ( y_{ H_I}, y_{t} ) } + \frac{1 }{ D^{b}_\gamma ( y_{A_I}, y_{t} ) }  \right]\,, \label{eq:Bga}
\end{align}
with
 \begin{align}
 D^{a}_\gamma (x, y)&= 1 - (1- x ) x_1  - y (1- x_1) x_2 \,, \nonumber \\
  D^{b}_\gamma (x, y) &= x_1 + x (1- x_1) - y  (1- x_1) x_2\,. 
 \end{align}
It can be seen that $C^{hc}_L$ and $B^{\gamma d}_{L}$, which arise from Fig.~\ref{fig:ttoqh}(c) and (d), are much smaller than other contributions because of  the suppression factor $y_t=m^2_t/m^2_{T}$.  To illustrate the smallness of $C^{hc}_L$ and $B^{\gamma d}_{L}$,  we take $m_{H_I}=65$ GeV, $m_{A_I, H^\pm_I}=100$ GeV, $m_{B_1, T(B)}=(1, 1.5)$ TeV, $c_\theta =0.78$, and $Y^{u(B')}_{Iq, I3}=2$; as a result,  $C^{hb}_{12R}\approx 3.2 \times 10^{-3}$, $C^{hc}_{L}\approx -9.6\times 10^{-6}$, $B^{\gamma a}_{1L} \approx -1.1 \times 10^{-3}$, and $B^{\gamma d}_{L}\approx 2.1\times 10^{-5}$.   Hence, the contributions from $C^{hc}_{L}$ and $B^{\gamma d}_{L}$ can be neglected.

 Using the obtained effective Wilson coefficients, the BRs for $t\to q (h, \gamma)$ can be estimated by the following relations
  \begin{align}
  Br(t\to q h) & =  \frac{m_t}{ 32 \pi  \Gamma_t }  \left(1-\frac{m^2_h}{m^2_t} \right)^2 \left( |C^h_L|^2 + |C^h_R|^2\right)\,, \nonumber \\
  Br(t\to q \gamma) &= \frac{\alpha m_t}{ 4  \Gamma_t }  \left( |B^\gamma_L|^2 + |B^\gamma_R|^2\right)\,, \label{eq:BRtqga}
  \end{align}
  where $\Gamma_t$ is the top-quark width. Since  the top-quark  decay is dominated by the $t\to Wb$ process, for the numerical estimation, we  take the next-to-leading order  SM  result for $\Gamma_t$, which is given as~\cite{Jezabek:1988iv,PDG}
   \begin{equation}
   \Gamma_t = \frac{G_F m^3_t}{8\pi \sqrt{2}} \left(1-\frac{m^2_W}{m^2_t} \right)^2 \left(1+ 2 \frac{m^2_W}{m^2_t} \right) \left[ 1-\frac{2 \alpha_s}{3\pi} \left( \frac{2\pi^2}{3} - \frac{5}{2}\right)\right]\,.
   \end{equation}

  \subsection{ $t\to q Z$ decay}

\begin{figure}[phtb]
\begin{center}
\includegraphics[scale=0.35]{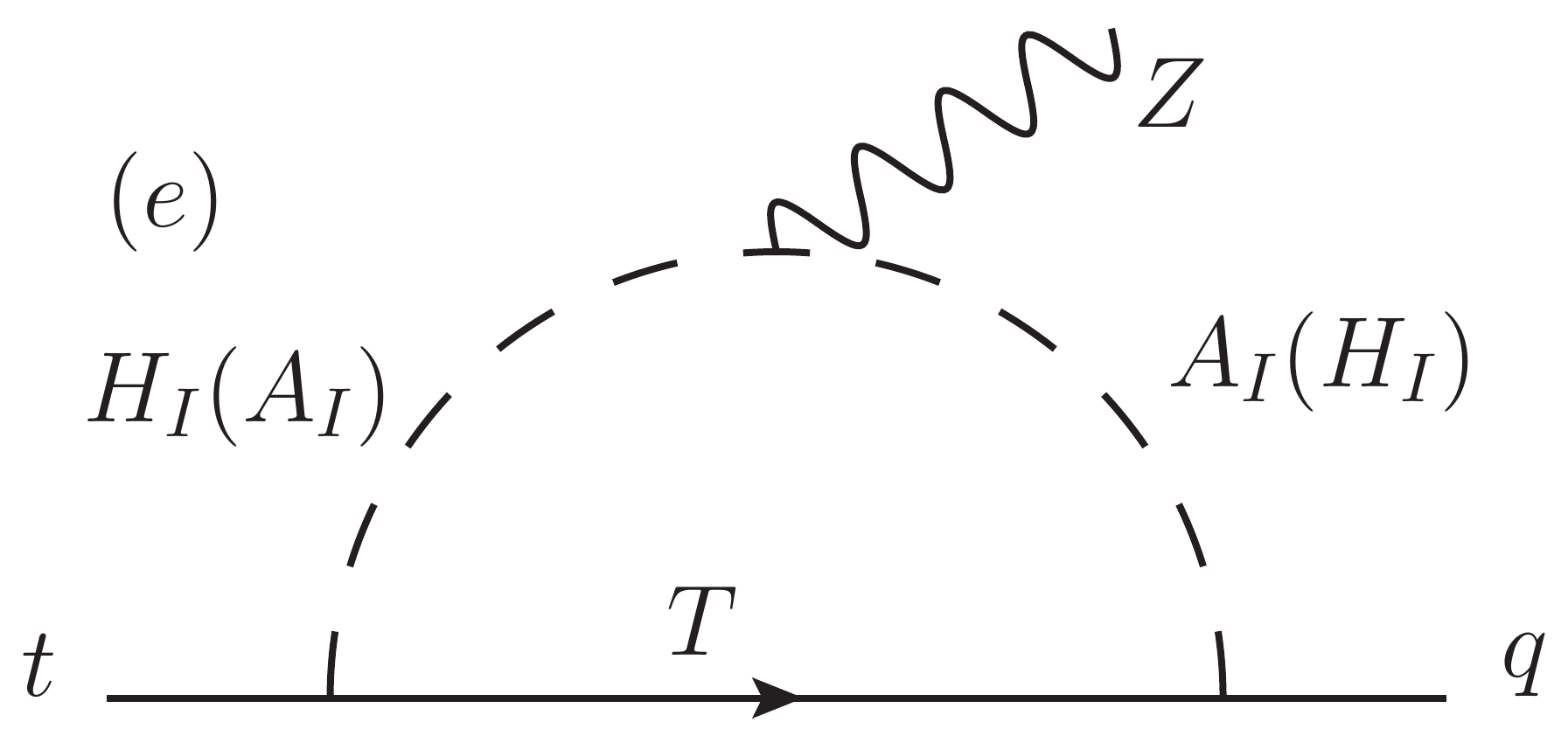}
 \caption{ Feynman diagram mediated by $H_I$ and $A_I$ for  the $t\to q Z$ process.}
\label{fig:ttoqZ}
\end{center}
\end{figure}

The $t\to q Z$ decay can arise from  Fig.~\ref{fig:ttoqh}(a), (b), and (d)  using $Z$ instead of  $h/\gamma$. Moreover,  the $t\to q Z$ decay can be produced via the $ZH_I A_I$ coupling, where the associated Feynman diagram is shown in Fig.~\ref{fig:ttoqZ}.  

 The decay amplitudes for the $t\to q Z$ decay can be formulated as
  \begin{align}
{\cal L}_{t\to q Z}  &=\bar u_q \gamma_\mu  \left( A^Z_L P_L + A^Z_R P_R\right ) t Z^\mu + \frac{1}{m_t} \bar u_q i\sigma_{\mu \nu} k^\nu \left( B^Z_L P_L + B^Z_R P_R   \right) t Z^\mu\,.
  \end{align}
 Because of the massive $Z$ boson, the $t\to q Z$ decay involves the vector currents in addition to the tensor currents. 
  Following the Feynman diagrams and using the couplings given in Eq.~(\ref{eq:Yukawa}) and Eqs.~(\ref{eq:VSS}-\ref{eq:VBB}), the effective Wilson coefficients  can be summarized as
 \begin{align}
 A^Z_{\chi} & = \sum_i \left( A^{Z a}_{i \chi} + \sum_j A^{Z b}_{ij \chi} \right) +A^{Zd}_\chi+ A^{Z e}_\chi\,, \nonumber \\ 
 B^Z_\chi & = \sum_{i} \left( B^{Z a}_{i\chi}  + \sum_j B^{Z b}_{ij\chi} \right)+ B^{Zd}_\chi +B^{Z e}_{\chi}\,.
 \end{align}
The contributions from each Feynman diagram are shown as
  \begin{align}
  A^{Za}_{iL}&= - \frac{g (c^2_{W} -s^2_W) }{2c_W (4\pi)^2}    \int^1_0 dx_1 \int^{x_1}_0 dx_2  \frac{1-x_1}{D^a_h(r_{i H^\pm_I}, r_{it}, r_{iZ} )} C^q_{iL} \left( C^t_{iL} r_{it} x_2  + C^t_{iR} \sqrt{r_{i t}} \right)\,, \nonumber \\
  B^{Za}_{iL} & = - \frac{g (c^2_{W} -s^2_W) }{2c_W (4\pi)^2}    \int^1_0 dx_1 \int^{x_1}_0 dx_2  \frac{1-x_1}{D^a_h(r_{i H^\pm_I}, r_{it}, r_{iZ} )} C^q_{iR} \left( C^t_{iR} r_{it} x_2  + C^t_{iL} \sqrt{r_{i t}} \right)\,, \nonumber \\
  X^{Zb}_{ij \chi} & = - \frac{g C^Z_{ij} } {c_W (4\pi)^2}  \int^1_0 dx_1 \int^{x_1}_0 dx_2 \frac{\tilde{X}^{Z}_{ij \chi}}{D^b_{h}(r_{j H^\pm_I}, r_{jt}, r_{ji}, r_{jZ})} \,, \nonumber 
  \end{align}
  \begin{align}
  A^{Zd}_{R} & = - \frac{g (1-2 Q_t s^2_W)} {4c_W (4\pi)^2} Y^{u}_{Iq} Y^u_{I3} \int^1_0 dx_1 \int^{x_1}_0 dx_2 \left(\frac{1+y_Z x_2 (x_1-x_2)}{D^b_{h}(y_{H_I}, y_{t}, 1, y_{Z})} + \frac{1+y_Z x_2 (x_1-x_2)}{D^b_{h}(y_{A_I}, y_{t}, 1, y_{Z})} \right)\,, \nonumber \\
   B^{Zd}_{L} & = \frac{g (1-2 Q_t s^2_W)} {4c_W (4\pi)^2} y_t Y^{u}_{Iq} Y^u_{I3} \int^1_0 dx_1 \int^{x_1}_0 dx_2 \left( \frac{x_2 (1-x_1)}{D^b_{h}(y_{H_I}, y_{t}, 1, y_{Z})}+  \frac{ x_2 (1-x_1)}{D^b_{h}(y_{A_I}, y_{t}, 1, y_{Z})} \right)\,, \nonumber \\
 A^{Ze}_{R} & = \frac{g  y_t}{4 c_W (4 \pi)^2}   Y^u_{Iq} Y^u_{I3} \int^1_0 dx_1 \int^{x_1}_0 dx_2   \left( \frac{x_2 (1-x_1)}{D^e_Z(y_{H_I}, y_{A_I}, y_{t}, y_{Z})} + \frac{x_2 (1-x_1)}{D^e_Z(y_{A_I}, y_{H_I}, y_{t}, y_{Z})}\right)\,,  \label{eq:tqZ}
  \end{align}
 $A^{Za}_{iR} = B^{Za}_{iL}$, $B^{Za}_{iR}= A^{Za}_{iL}$, $A^{Zd}_{L}=B^{Zd}_{R}=0$, and $B^{Ze}_{L} =A^{Ze}_{R}$ with  
  \begin{align}
  \tilde A^Z_{ij L} & = C^q_{jL} \left[ C^t_{iL} \left( \sqrt{r_{ji}} + r_{jZ} x_2 (x_1-x_2) \right) + C^t_{iR} \sqrt{r_{jt}} (1-x_1) \right]\,, \nonumber \\
  \tilde A^Z_{ij R} & = C^q_{jR} \left[ C^t_{iR} \left( \sqrt{r_{ji}} + r_{jZ} x_2 (x_1-x_2) \right) + C^t_{iL} \sqrt{r_{jt}} (1-x_1) \right]\,, \nonumber \\
   \tilde B^Z_{ij L} & = -C^q_{jR} \left[  C^t_{iR} r_{jt} x_2 (1-x_1) + C^t_{iL} \left( \sqrt{r_{jt} r_{ji} } x_2 + \sqrt{r_{jt}} (x_1 -x_2)\right) \right]\,, \nonumber \\
   \tilde B^Z_{ij R} & = -C^q_{jL} \left[  C^t_{iL} r_{jt} x_2 (1-x_1) + C^t_{iR} \left( \sqrt{r_{jt} r_{ji} } x_2 + \sqrt{r_{jt}} (x_1 -x_2)\right) \right]\,, \nonumber \\
   D^e_Z(w,x,y,z) & = 1 + (1-x) x_1 -(y-w-x) x_2 + z x^2_2 + y x_1 x_2\,,
  \end{align}
  where $X^{Zb}_{ij\chi}$ and $\tilde{X}^{Z}_{ij\chi}$, which involve the $Z$ flavor-changing couplings,  are the contributions from Fig.~\ref{fig:ttoqh}(b).    From Eq.~(\ref{eq:tqZ}), it can be seen that $A^{Ze}_R$ from Fig.~\ref{fig:ttoqZ} is suppressed by $m^2_t/m^2_T$. According to earlier analysis, the contribution is small and can be neglected. Moreover, the $Z$-penguin induced from  Fig.~\ref{fig:ttoqh}(d) is comparable with that induced from Fig.~\ref{fig:ttoqh}(b). 
  Using the obtained $A^Z_\chi$ and $B^Z_\chi$, the BR for the $t\to q Z$ decay can be written as
   \begin{align}
   Br(t\to q Z) & =\frac{1}{\Gamma_t} \frac{G_F m_t^3}{4\sqrt{2} \pi}  \frac{c^2_W}{g^2} \left( 1 -\frac{m^2_Z}{m^2_t}\right)^2 \sum_{\chi=L,R} \left[ |A^Z_\chi + B^Z_{\chi'} |^2 \left( 1+ \frac{ 2m^2_Z}{m^2_t}\right) \right. \nonumber \\
   &  \left. + \left( |B^Z_{\chi'}|^2 -2 Re(A^Z_\chi + B^Z_{\chi'}) B^{Z*}_{\chi'}) \right)\left( 1- \frac{ m^2_Z}{m^2_t}\right)\right]\,, \label{eq:BRtqZ}
   \end{align}
where $B^Z_{\chi'}$ denotes that when $A^Z_{\chi}=A^{Z}_{L(R)}$,   $B^Z_{\chi'}=B^Z_{R(L)}$.   

\section{Constraints} \label{sec:constraints}

In this section, we discuss the possible constraints from the flavor physics, DM data, and  oblique parameters. 

\subsection{$b\to q' \gamma$ and $\Delta B=2$}

The similar Feynman diagrams for $t\to q \gamma$ can be applied to the $B\to X_{q'} \gamma$ ($q'=d,s$) decays, where the current measurements are $BR(B\to X_s \gamma) = (3.49 \pm 0.19)\times 10^{-4}$ and $BR(B\to X_d \gamma) = (9.2 \pm 3.0)\times 10^{-6}$~\cite{PDG}. If there are large effects contributing to the radiative $B$ decays, then the current data may provide a serious bound. 
 From Eq.~(\ref{eq:Yukawa}), it can be seen that  the involving Yukawa couplings for the $b\to q' \gamma$ decay are ${\bf Y}^{B'}_I$ and ${\bf Y}^d_I$.  
 To understand the influence from ${\bf Y}^{B'}_I$ and ${\bf Y}^{d}_I$, we write the effective interaction for $b\to q' \gamma$ from the new physics as
\begin{align}
{\cal L}_{b\to q' \gamma} &=  \frac{G_F V_{tb} V^*_{tq'} }{\sqrt{2}} \left( C^{\rm NP}_{7R} {\cal O}_{7R}  + C^{\rm NP}_{7L} {\cal O}_{7L} \right)\,, \nonumber \\
 {\cal O}_{7\chi} & = \frac{e m_b}{4 \pi^2} \overline{d_{q'}} \sigma_{\mu \nu} P_\chi b F^{\mu \nu} \,.
\end{align}
 The  $C^{\rm NP}_{7R,7L}$ mediated by $S_I$ and $H^\pm_I$ can be expressed as
\begin{align}
C^{\rm NP,q'}_{7R} & = -\frac{\sqrt{2} Y^{B'}_{Iq'} Y^{B'}_{I3} }{G_F V_{tb} V^*_{tq'} } \sum^2_{i=1} \frac{ Q_b \zeta^2_i }{16 m^2_{B_i}} \left( J(r_{iH_I}) + J(r_{i A_I})\right)\,, \nonumber \\
C^{\rm NP,q'}_{7L} & = - \frac{\sqrt{2} Y^{d}_{Iq'} Y^{d}_{I3} }{G_F V_{tb} V^*_{tq'} } \left[ \frac{ 1 }{16 m^2_{T}} \left( J' (y_{H^\pm_I})  Q_t J(y_{H^\pm_I})\right) \right.\nonumber \\ 
& \left. - \sum^2_{i=1} \frac{ Q_b \xi^2_i }{16 m^2_{B_i}} \left( J(r_{iH_I}) + J(r_{i A_I})\right) \right]\,, \nonumber \\
 J(a) & =  \frac{1-5a -2a^2}{6(1-a)^3} - \frac{a^2 \ln a}{2 (1-a)^4}\,, \nonumber \\
 J'(a) &= \frac{2+5a-a^2}{12 (1-a)^3} + \frac{a \ln a}{2(1-a)^4}\,,
\end{align}
with $\zeta_1=c_\theta$, $\zeta_2=s_\theta$, $\xi_1=-s_\theta$, and $\xi_2=c_\theta$. 
 Using $|V_{ts}|=0.04$, $|V_{td}|=0.0088$, $s_\theta=1/\sqrt{2}$, ${\bf Y}^{B'}_{I}= {\bf Y}^d_{I}= 2$, and $m_{B_1, B_2, T}=(1, 1.5, 1.4)$ TeV, we obtain $C^{\rm NP,s}_{7R,7L}\sim (-0.051,\, 0.033)$ and $C^{\rm NP,d}_{7R,7L}\sim (0.23,\, -0.15)$,  where   the SM result is $C^{\rm SM}_{7R,7L}\sim (-0.3,\, 0)$~\cite{Buchalla:1995vs}. It can be seen that due to $|V_{td}|/|V_{ts}|\sim 0.22$, the values of $C^{\rm NP,d}_{7\chi}$ are much larger than those of $C^{\rm NP,s}_{7\chi}$. Although the value of $C^{\rm NP,d}_{7R}$ is close to that of $C^{\rm SM}_{7R}$, the contribution from ${\cal O}_{7R}$ operator can be diminished due to the opposite sign in $C^{\rm NP,d}_{7R}$ and $C^{\rm SM}_{7R}$; thus, the dominant contribution to $B\to X_d \gamma$ is from ${\cal O}_{7L}$.  Since $|C^{\rm NP,d}_{7L}|$ is less than $|C^{\rm SM}_{7R}|$, ${\bf Y}^{B'(d)}_{I} \sim 2$ are still allowed when the constraint from the $B\to X_d \gamma$ process is taken into account.  Hence, $B\to X_{q'} \gamma$ do not provide  severe constraints on the parameters $Y^{B'}_{Iq'}$, which are related to the $t\to q (h, V)$ decays.  

 In addition to the radiative $b$ decays, the Yukawa couplings ${\bf Y}^{B'}_I$ and ${\bf Y}^d_{I}$  can also  contribute to the $\Delta B=2$ process through box diagrams mediated by $H_I$, $A_I$, and $H^\pm_I$. Since the Yukawa couplings to $S_I$ involves left-handed and right-handed couplings, for simplicity, we write the Yukawa couplings as
  \begin{align}
  {\cal L}_Y \supset - \eta_{S_I} \bar B_i ( C^i_{L f} P_L + C^i_{Rf} P_R) f S_I + H.c.\,,
  \end{align}
 where $\eta_{H_I}=1$, $\eta_{A_I}=-\sqrt{-1}$,  $C^{i}_{Lf}=\zeta_i Y^{B'}_{If}/\sqrt{2}$, and $C^{i}_{Lf}=-\xi_i  Y^{d}_{If}/\sqrt{2}$. In order to simplify the expression and show the possible destruction  between ${\bf Y}^{B'}_{I}$ and ${\bf Y}^{d}_{I}$, we take $m_{F}\equiv m_T\sim m_{B_1} \sim m_{B_2}$ and neglect  the small ratios  $m^2_{S_I, H^\pm_I}/m^2_{B_1}$; as a result,  the effective Lagrangian for $\Delta B=2$ can be written as
  \begin{align}
  {\cal L}_{\Delta B=2} & \sim  - \frac{1}{2 (4\pi)^2} \sum^2_{i,j=1}  \frac{1}{m^2_{B_i}} \left[ C^i_{Lq'} C^i_{L3} C^j_{Lq'} C^j_{L3} (\bar q' \gamma_\mu P_L b)^2  + C^i_{Rq'} C^i_{R3} C^j_{Rq'} C^j_{R3} (\bar q' \gamma_\mu P_R b)^2 \right. \nonumber \\ 
  & \left.  + \left( C^i_{Lq'} C^i_{L3} C^j_{Rq'} C^j_{R3} + R\leftrightarrow L \right) (\bar q' \gamma_\mu P_L b) (\bar q' \gamma^\mu P_R b) \right] -  \frac{Y^d_{Iq'} Y^d_{I3}}{8 (4\pi)^2 m^2_T} (\bar q' \gamma_\mu P_R b)^2\,.
  \end{align}
 Using $\sum_i \zeta^2_i =\sum_i \xi^2_i = 1$ and the hadronic matrix elements of $B_{q'}$ for various effective operators that were obtained in Ref.~\cite{Buras:2001ra}, the transition matrix element for $B_{q'}$ mixing is obtained as
 \begin{align}
 \langle B_{q'} | H_{\rm eff}| \bar B_{q'} \rangle & \sim  \frac{f^2_{B_{q'}} m_{B_{q'}} }{6 (4\pi)^2 m^2_{F}} \left[ \left( (Y^{B'}_{Iq'} Y^{B'}_{I3})^2  +\frac{5}{4}  (Y^{d}_{Iq'} Y^{d}_{I3})^2 \right) P^{\rm VLL}_1 \right. \nonumber \\
 & \left. + 2 Y^{B'}_{Iq'} Y^{B'}_{I3} Y^{d}_{Iq'} Y^{d}_{I3}\,  P^{LR}_1\right]\,, \label{eq:Bq_mixing}
 \end{align}
 where $P^{VLL}_1$ and $P^{LR}_1$ are the nonperturbative  hadronic effects of $(\bar q' \gamma_\mu P_{R(L)} b)^2$ and $(\bar q' \gamma_\mu P_L b) (\bar q' \gamma^\mu P_R b)$, respectively, and their values are $P^{VLL}_1\approx 0.84$ and $P^{LR}_{1}\approx -1.62$~\cite{Buras:2001ra}.   It can be seen that because $P^{VLL}_1$ and $P^{LR}_1$ are opposite in sign, the contributions from ${\bf Y}^{B'}_{I}$ to $\Delta B=2$ can be diminished by those from ${\bf Y}^d_{I}$. To numerically exhibit  the cancellation, we take $\hat w=Y^{d}_{Iq'} Y^{d}_{I3}$ as a variable, fix $Y^{B'}_{Iq'} =Y^{B'}_{I3}=2$, and turn  Eq. (\ref{eq:Bq_mixing}) to be a quadratic equation as
 \begin{equation}
\frac{5}{4} P^{VLL}_1 \hat w^2 + 8 P^{RL}_1 \hat w + 16 P^{VLL}_1=0  \,. \label{eq:w}
 \end{equation}
 By solving the quadratic equation, the values of $Y^d_{Iq'} Y^d_{I3}$, which lead to small $B_{q'}$ mixing, can then be found. Indeed, two solutions to  Eq.~(\ref{eq:w}) exist and are obtained as: $\hat w\approx 1.143$ and $\hat w\approx 11.2$.  Based on the analysis, it is known that the strict constraint from the $\Delta B=2$ process can be avoided when the left-handed and right-handed current couplings are simultaneously considered.

\subsection{ Oblique parameters} 

 Since  the $Z_2$-odd quark $B$ in doublet $Q_{4}$ mixes with the singlet $B'$, the mixing effect leads to the mass difference between $T$ and $B$.  In Eq.~(\ref{eq:massBBp}), it can be seen that the mass splitting within the vector-like quark doublet can be expressed as $\delta m_{Q_4} =  |m_T - m_{B_2}|$ and is dictated by $v y_{B'}/\sqrt{2}$.
This mass splitting contributes to the electroweak oblique parameters, where the current measurements with $U=0$ are given as~\cite{PDG}
\begin{equation}
S = 0.02 \pm 0.07\,, ~ T =0.06 \pm 0.06\,. \label{eq:ST_data}
 \end{equation}
Therefore, the precision measurements of electroweak oblique parameters~\cite{Peskin:1991sw} may constrain $y_{B'}$ or  the mixing angle $\theta$.  Take the constraints into account, following the results in~~\cite{Lavoura:1992np,Chen:2017hak}, we write the oblique correction to the $T$ parameter as
 \begin{align}
 \Delta T_{Q_4} & =\frac{N_c}{8 \pi s^2_W c^2_W} \left[ s^2_\theta \Theta_{+-}(z_T, z_{B_1}) + c^2_\theta \Theta_{+-}(z_T, z_{B_2}) -s^2_\theta c^2_\theta \Theta_{+-}(z_{B_1}, z_{B_2})\right] \,,  \label{eq:Q_T}
 \end{align}
where $N_c=3$ is the color number, $z_f=m^2_f/m^2_Z$,  and  $\Theta_{+-}(a,b) = \theta_{+}(a,b) + \theta_{-} (a,b)$, with
  \begin{align}
  \theta_+ (a,b) & =a+b-\frac{2 a b}{a-b} \ln\left( \frac{a}{b}\right)\,, \nonumber \\
  \theta_{-}(a,b) & = 2\sqrt{a b} \left( \frac{a+b}{a-b}  \ln\left( \frac{a}{b}\right)-2 \right)\,.
  \end{align}
  When $y_{B'} =0$, it can be seen that $s_\theta =0$ and $m_T=m_{B_2}=m_{4Q}$. Due to $\theta_{\pm}(a,a)=0$, we obtain $\Delta T=0$. 
  
The correction to the $S$ parameter can be expressed as
\begin{align}
 \Delta S_{Q_4} & =\frac{N_c}{\pi} \left[ s^2_\theta \Psi_{+-}(z_T, z_{B_1}) + c^2_\theta \Psi_{+-}(z_T, z_{B_2}) -s^2_\theta c^2_\theta \chi_{+-}(z_{B_1}, z_{B_2})\right] \,, 
 \end{align}
where $\Psi_{+-}(a,b)=\Psi_+(a,b) + \Psi_{-}(a,b)$ and $\chi_{+-}(a,b)=\chi_+(a,b) + \chi_{-}(a,b)$, with
 \begin{align}
\Psi(a,b)&= \frac{1}{3} - \frac{1}{9} \ln\left( \frac{a}{b}\right)\,, ~ \Psi_{-}(a,b) = - \frac{a+b}{6\sqrt{a b}}\,, \nonumber \\
\chi_{+}(a,b) & = \frac{5(a^2 + b^2) -22 a b}{9(a-b)^2 } + \frac{3ab(a+b) -a^3 -b^3}{3(a-b)^3} \ln \left( \frac{a}{b}\right)\,, \nonumber \\
\chi_{-}(a,b) & = - \sqrt{ab} \left[  \frac{a+b}{6a b} - \frac{a+b}{(a-b)^2} + \frac{2 a b}{(a-b)^3}  \ln \left( \frac{a}{b}\right)\right] \,.
 \end{align}
 Similar to $\Delta T_{Q_4}$, when $y_{B'}=0$, due to $\Psi_{+-}(a,a)=0$ and $\chi_{\pm}(a,a)=0$, we obtain $\Delta S_{Q_4}=0$. With $c_\theta=0.8$,   $\Delta S_{Q_4} \sim 0.01$, which is much smaller than $\Delta T_{Q_4}$. Thus, we only take the $T$ parameter as the potential constraint.
 
 The mass splittings among $H^\pm_I$, $H_I$, and $A_I$ also contribute to the $T$ parameter. Following the results shown in~\cite{Barbieri:2006dq}, the correction of the inert Higgs doublet to the $T$ parameter is expressed as
  \begin{align}
  \Delta T_{\Phi_I} = \frac{1}{16 \pi s^2_W c^2_W} \left( \theta_{+}(z_{H^\pm_I}, z_{H_I}) +\theta_{+}(z_{H^\pm_I}, z_{A_I}) - \theta_{+}(z_{H_I}, z_{A_I})) \right)\,. \label{eq:S_T}
  \end{align}
  
 
\subsection{Higgs production and $h\to \gamma\gamma$}

 From Eqs.~(\ref{eq:hBB}) and  (\ref{eq:hHIHI}), the SM Higgs has extra couplings to  $H^\pm_I$ and $B_i$, where the former can induce the $h\gamma \gamma$ effective coupling, and  the latter can generate $h\gamma \gamma$ and $hgg$ effective couplings. With the precision measurements for the  Higgs production and Higgs decay to diphoton, the new physics effect could be strictly bounded. To show the new physics effect, the signal strength for $pp\to h \to \gamma \gamma$ is defined as
 \begin{align}
 \mu_{\gamma\gamma} = \frac{\sigma(pp\to h)}{\sigma(pp\to h)^{\rm SM}} \frac{BR(h\to \gamma \gamma)} {BR(h\to \gamma\gamma)^{\rm SM}} \,,
 \end{align}
where the  measurements  from ATLAS and CMS at $\sqrt{s}=13$ TeV are  given as $1.10\pm 0.10$ using $139$ fb$^{-1}$ ~\cite{ATLAS:2020pvn} and $1.03^{+0.11}_{-0.09}$ using $137$ fb$^{-1}$~\cite{CMS:2020omd}, respectively.

  \begin{figure}[phtb]
\begin{center}
\includegraphics[scale=0.55]{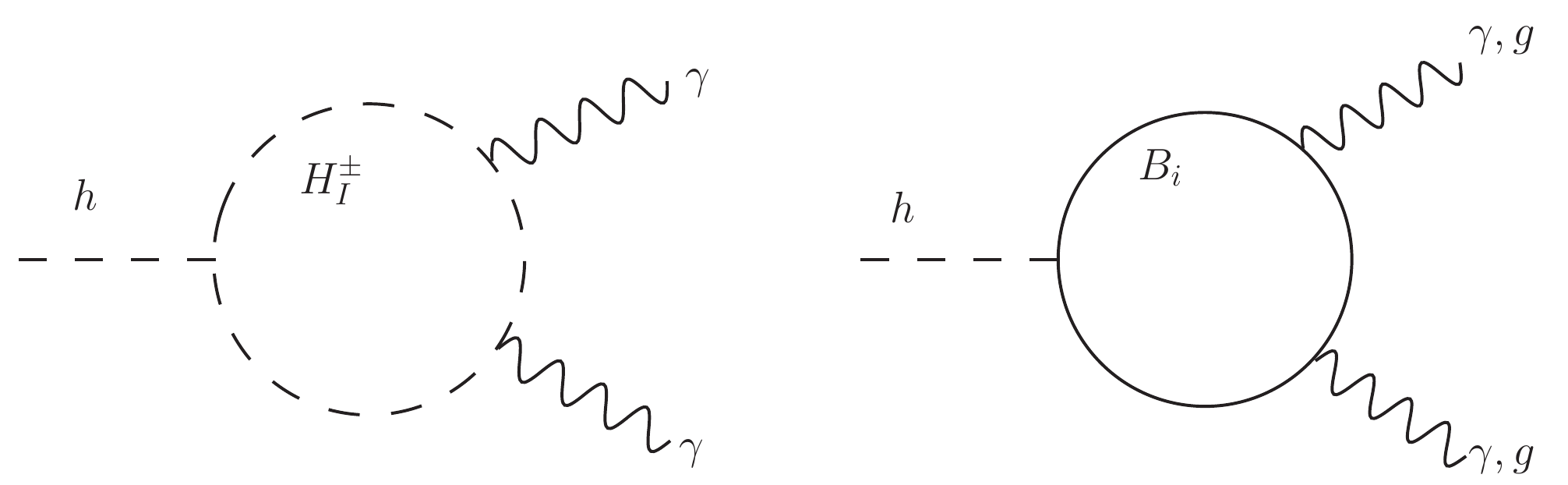}
 \caption{ Feynman diagram for $h\to (\gamma\gamma, gg)$ induced by $H^\pm_I$ and $B_i$. }
\label{fig:hVV}
\end{center}
\end{figure}

The loop-induced effective interactions for $h\gamma\gamma$ and $hgg$ can be parameterized as
\begin{align}
{\cal L}_{hVV} & = \frac{\alpha}{4\pi} \frac{a_{\gamma\gamma}}{m_h}  h F_{\mu \nu} F^{\mu \nu} + \frac{\alpha_s}{4\pi} \frac{a_{gg}}{m_h} h G^{a}_{\mu\nu} G^{a\mu\nu}\,,
\end{align} 
where the Feynman diagrams mediated by $H^\pm_I$ and $B_i$ are shown in Fig.~\ref{fig:hVV}.  The resulting $a_{\gamma\gamma}$ and $a_{gg}$ in the model  are obtained as
\begin{align}
a_{\gamma \gamma} & = \frac{g m_h }{2m_W} \left( a^{\rm SM}_{2\gamma} +  \frac{\lambda_3 v m_W}{g m^2_{H^\pm_I}} F_0(\tau_{H^\pm_I}) +  N_c Q^2_b \sum_i  \frac{ y^h_{ii} v}{\sqrt{2} m_t} F_{1/2} (\tau_{B_i}) \right)\,, \nonumber \\
a_{gg} & = \frac{g m_h }{4 m_W} \left( a^{\rm SM}_{2g} + \sum_i  \frac{ y^h_{ii} v}{\sqrt{2} m_t} F_{1/2} (\tau_{B_i}) \right)\,
\end{align}
where $a^{\rm SM}_{2\gamma}\approx 6.51 - i 0.02$ and $a^{\rm SM}_{2g}\approx -0.69$ are the SM results; $y^{h}_{11}=-s_{2\theta} y_{B'} $ and $y^h_{22}=s_{2\theta} y_{B'}$, and the functions $F_{0}$ and $F_{1/2}$ are given as
 \begin{align}
 F_0 (\tau) & =  \tau (1 - \tau f(\tau) )\,, \nonumber \\
 F_{1/2}(\tau) & = -2 \tau (1+(1-\tau) f(\tau))\,,
 \end{align}
 with $\tau =4 m^2_f/m^2_h$ and $f(\tau)= (\arcsin(1/\sqrt{\tau}))^2$. Considering that we focus on the case with $2 m_{H^\pm_I}, 2m_{B_i} > m_h$,  the on-shell condition in the loop propagators is not available. 

 When $m_{B_i} \gg m_h$, $F_{1/2}(\tau)\to -4/3$. Because $y^h_{11}=-y^h_{22}$,  the contributions  to the $hgg$ effective coupling from the $B_{1}$ and $B_{2}$ quarks are canceled each other; that is, the Higgs production through the gluon-gluon fusion is the same as the SM. Thus, the signal strength for $pp\to h \to \gamma\gamma$ can be simplified as
 \begin{equation}
 \mu_{\gamma\gamma}  \approx \frac{BR(h\to \gamma \gamma)} {BR(h\to \gamma\gamma)^{\rm SM}} \approx \left| 1 + \frac{\lambda_3 v m_W}{ g m^2_{H_I} a^{\rm SM}_{2\gamma} } F_{0} (\tau_{H^\pm_I}) \right|^2\,, \label{eq:mu2gamma_ICH}
 \end{equation}
where the new physics on the Higgs width $\Gamma_h$ is assumed to be small and neglected in $\mu_{\gamma\gamma}$. To suppress the invisible Higgs decay $h\to S_I S_I$ and to have  $\Gamma_h \approx \Gamma^{\rm SM}_h$, we simply take $m_h < 2 m_{S_I}$ in the model. Using $m_{H^\pm_I}=100$ GeV, the $H^\pm_I$ effect on $\mu_{\gamma\gamma}$ can be estimated as $-0.197 \lambda_3$. If we take the allowed range of $\mu_{\gamma\gamma}$ to be $0.8 < \mu_{\gamma\gamma}< 1.2$,  then $\lambda_3$ is limited to be $\lambda_3 < 0.5$.  Hence,  the $\lambda_3$ parameter can be bounded by the $h\to \gamma\gamma$ measurement.

\subsection{DM direct detection}

In the inert Higgs doublet model, although there is no $ZH_IH_I$ coupling at the tree level,   the nonvanizing $ZH_I A_I$ coupling will contribute to the DM-nucleon scattering. To satisfy the DM direct detection experiments, the $Z$-mediated $H_I n\to A_I n$ process has to be suppressed, where $n$ denotes the nucleon. The process can be kinematically forbidden by requiring $m_{A_I} -m_{H_I} $ to be larger than the kinetic energy of the DM,  where the typical energy is tens of keV. In the study, we take $m_{A_I} - m_{H_I} > 1$ GeV. 

 The spin-independent (SI) DM-nucleon scattering can occur via the trilinear coupling $hH_I H_I$ shown in Eq.~(\ref{eq:hHIHI}). The Higgs-mediated cross-section can be formulated as~\cite{Barbieri:2006dq}
  \begin{align}
  \sigma^{\rm SI}_{h} = \frac{\mu^2_{H_I n}}{4\pi} \left|  \frac{\lambda_L }{m_{H_I} m^2_h} \right|^2 f^2_n m^2_n\,,
  \end{align}
 where  $\mu_{H_I n}=m_{H_I} m_n/(m_{H_I} + m_n)$ is the DM-nucleon reduced mass, and $f_n\approx 0.3$ is the nucleon matrix element. With $m_n=0.94$ GeV and $m_{H_I}=70$ GeV, we have
 \begin{equation}
 \sigma^{\rm SI}_{h} \approx 2.0 \times 10^{-42} (\lambda_3 - |\lambda_4| - |\lambda_5|)^2 ~ \text{cm$^2$}\,. 
 \end{equation}
 To satisfy the XENON1T upper limit of $7\times 10^{-47}$ cm$^2$~\cite{Aprile:2018dbl},  $\lambda_3 - |\lambda_4| -|\lambda_5|< 5 \times 10^{-3}$ is required; that is, $\lambda_3 \sim |\lambda_4| + |\lambda_5|$. As mentioned earlier, $\lambda_3 $ is bounded by $\mu_{\gamma\gamma}$. Therefore, the magnitude $|\lambda_{4}|$ can be bounded by the DM direct detection. With the mass ordering of $m_{H^\pm_I} > m_{A_I} > m_{H_I}$, we obtain  $|\lambda_5| < |\lambda_4|$.

\section{ Numerical analysis and discussions}\label{sec:num}

Many free parameters are involved in the phenomenological analysis, such as Yukawa couplings ${\bf Y}^{u}_{I}$ and ${\bf Y}^{B'}_{I }$, the inert scalar masses $m_{H_I, A_I, H^\pm_I}$, the $Z_2$-odd quark masses $m_{B_1, B_2, T}$, and the parameter $\lambda_3$ in the scalar potential. Before discussing their influence on  the rare top decays, we first determine the allowed ranges for the free parameters and then use the constrained parameter values to  analyze the implications on the rare top decays. For the numerical analysis, the current experimental upper limits are taken as~\cite{PDG}
 \begin{align}
 BR(t\to q \gamma) & <  1.8 \times 10^{-4}\,, \nonumber \\
 BR(t \to q Z) & < 5 \times  10^{-4}\,, \nonumber \\
 BR(t\to u h) & < 1.2 \times 10^{-3}\,, \nonumber \\
 BR(t\to c h) & < 1.1 \times 10^{-3}\,. \label{eq:exp_upper}
 \end{align}
  
\subsection{Parameter choices and constraints}

As stated earlier, the radiative $B$ decay cannot severely bound the Yukawa couplings. Thus, we employ the perturbative unitarity constraint, and the upper limits of the Yukawa couplings are required to be $|{\bf Y}^{u(B')}_{I}| < 2\sqrt{2\pi}$~\cite{Castillo:2013uda}.  To obtain the mass upper limit  of the $Z_2$-odd quarks,   we apply the similar constraints for the stop and sbottom with the $R$-parity conserving supersymmetry, where using  the data with an integrated luminosity of 139 fb$^{-1}$ at $\sqrt{s}=13$ TeV~\cite{ATLAS:2021hza}, the mass below $1$ TeV has been excluded by ATLAS when the neutralino mass is below 100 GeV. 
Thus, for the parameter scan, we assume $m_{B_1} < m_{B_2}$ and take the mass regions for $m_{B_{1,2}}$ and $m_T$ to be
 \begin{equation}
 m_{B_1} \in (1000, 1200)~\text{GeV}\,,~m_{B_2, T}\in (1000,2000)~\text{GeV}\,. \label{eq:Qmass_region}
 \end{equation}
 If $H_I$ is the DM candidate in the inert Higgs model,  then $m_{H_I}\sim m_W$ can fit the observed DM relic abundance~\cite{Barbieri:2006dq}.  It has been studied that the direct bounds from colliders on $m_{H_I, A_I}$ are not strict, and the bound on $m_{H^\pm_I}$, which converts from the SUSY search at LEP, is $m_{H^\pm_I}> 70 -90$ GeV~\cite{Pierce:2007ut,Lundstrom:2008ai,Merchand:2019bod,Belanger:2021lwd}.  Therefore, the mass regions of $H_I$, $A_I$, and $H^\pm_I$ are taken as
 \begin{equation}
 m_{H_I, A_I, H^\pm_I} \in (70, 120)~\text{GeV}\,. \label{eq:smass_region}
 \end{equation}

\begin{figure}[phtb]
\begin{center}
\includegraphics[scale=0.35]{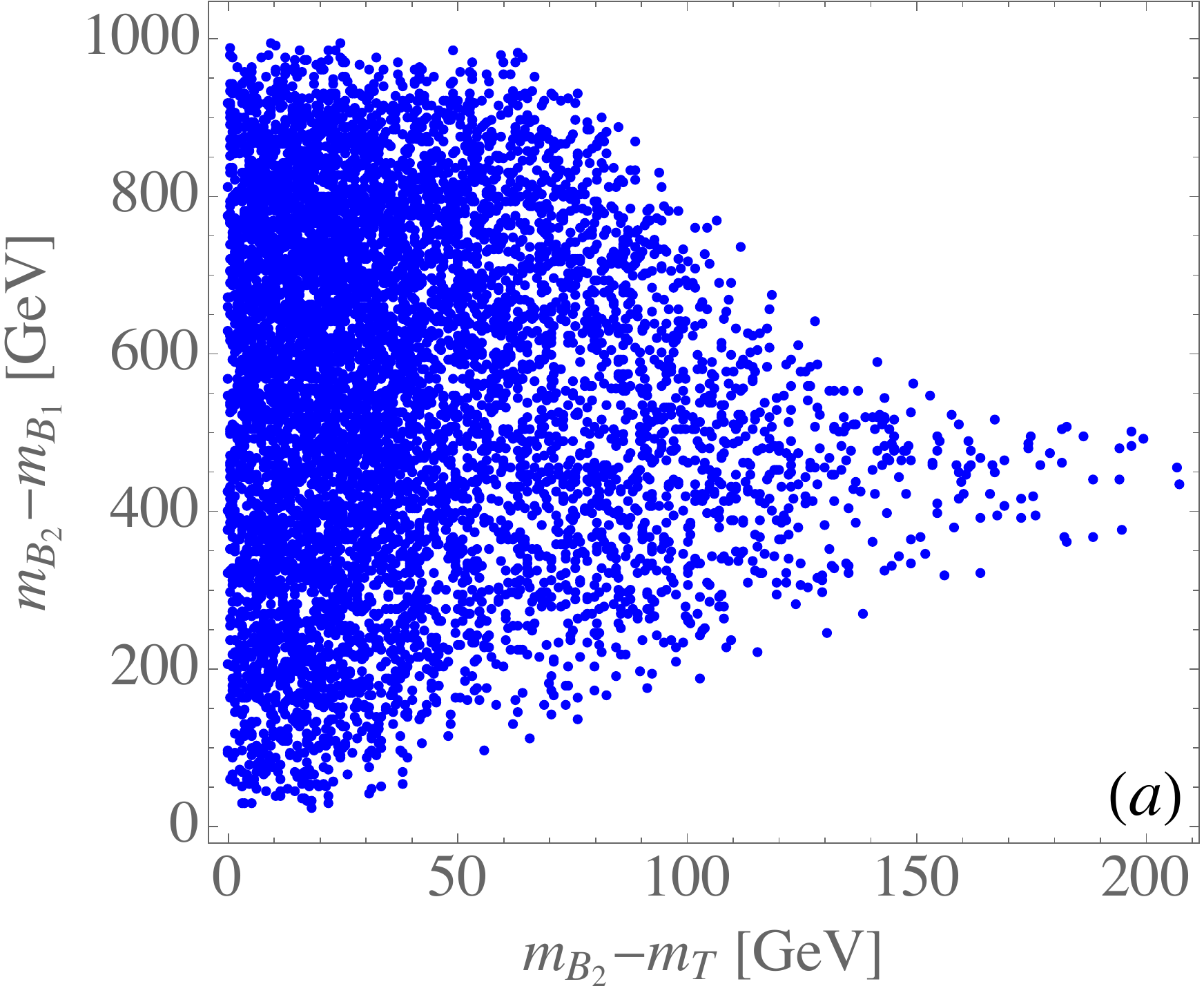}
\includegraphics[scale=0.35]{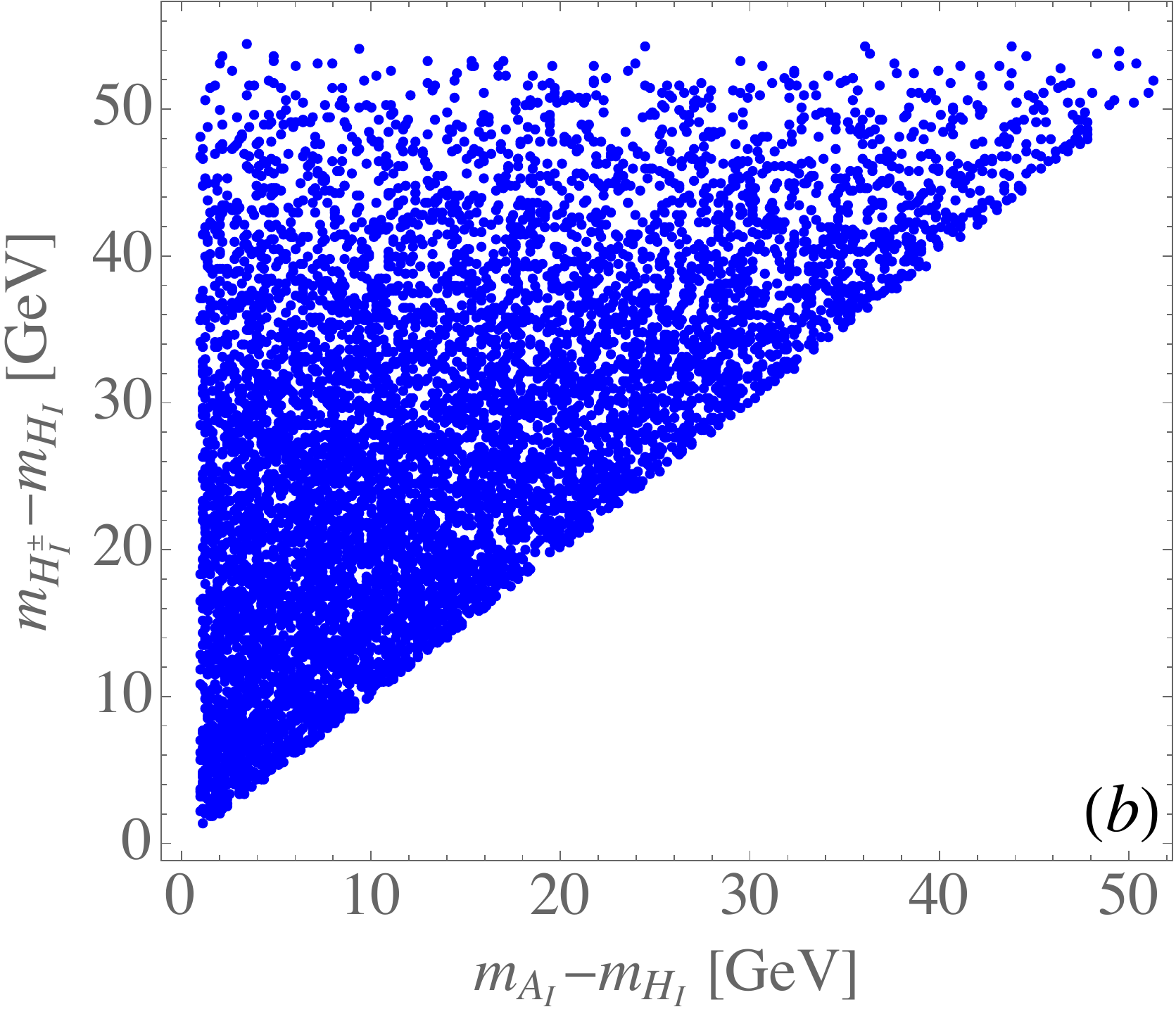}
\includegraphics[scale=0.35]{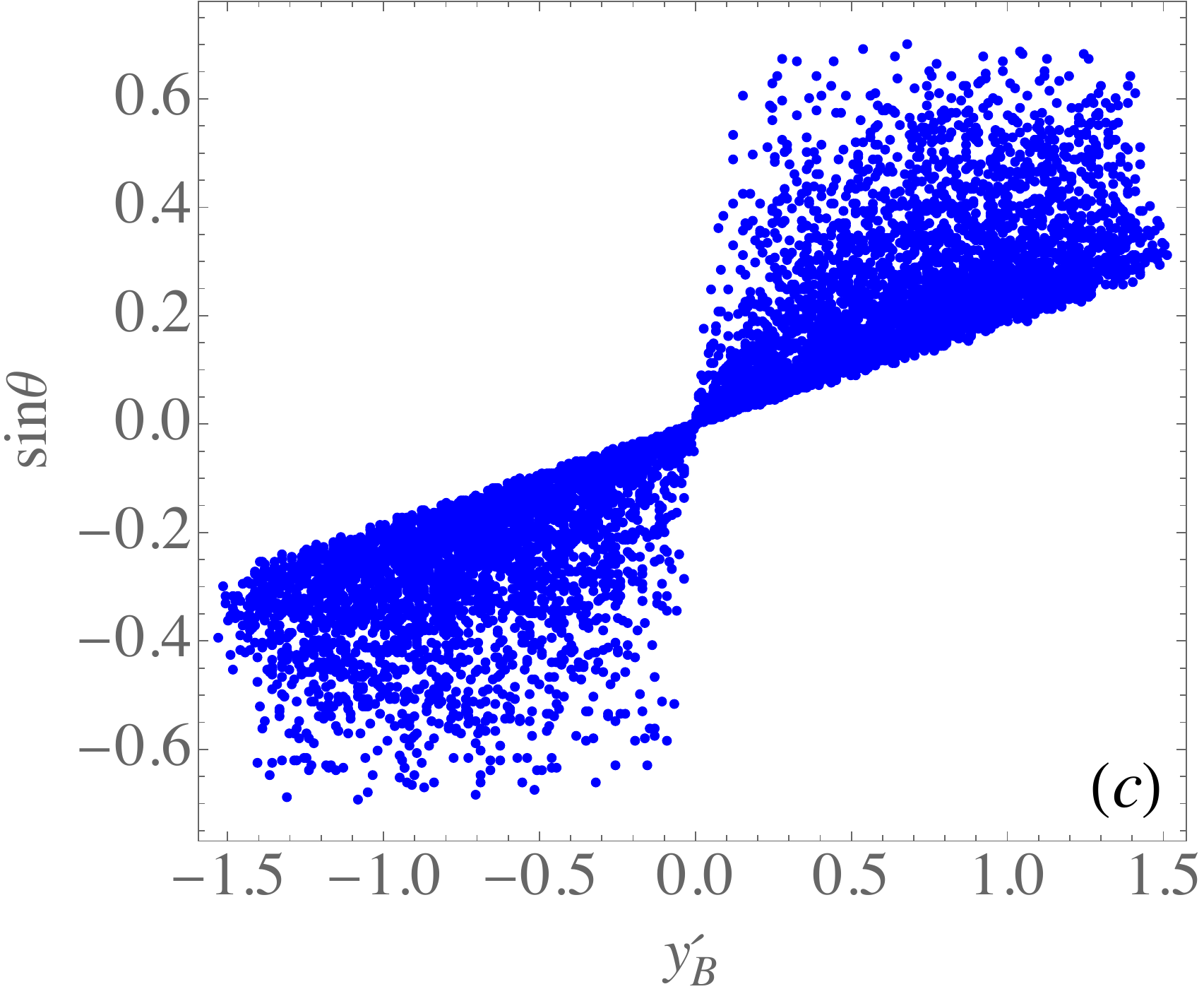}
\includegraphics[scale=0.35]{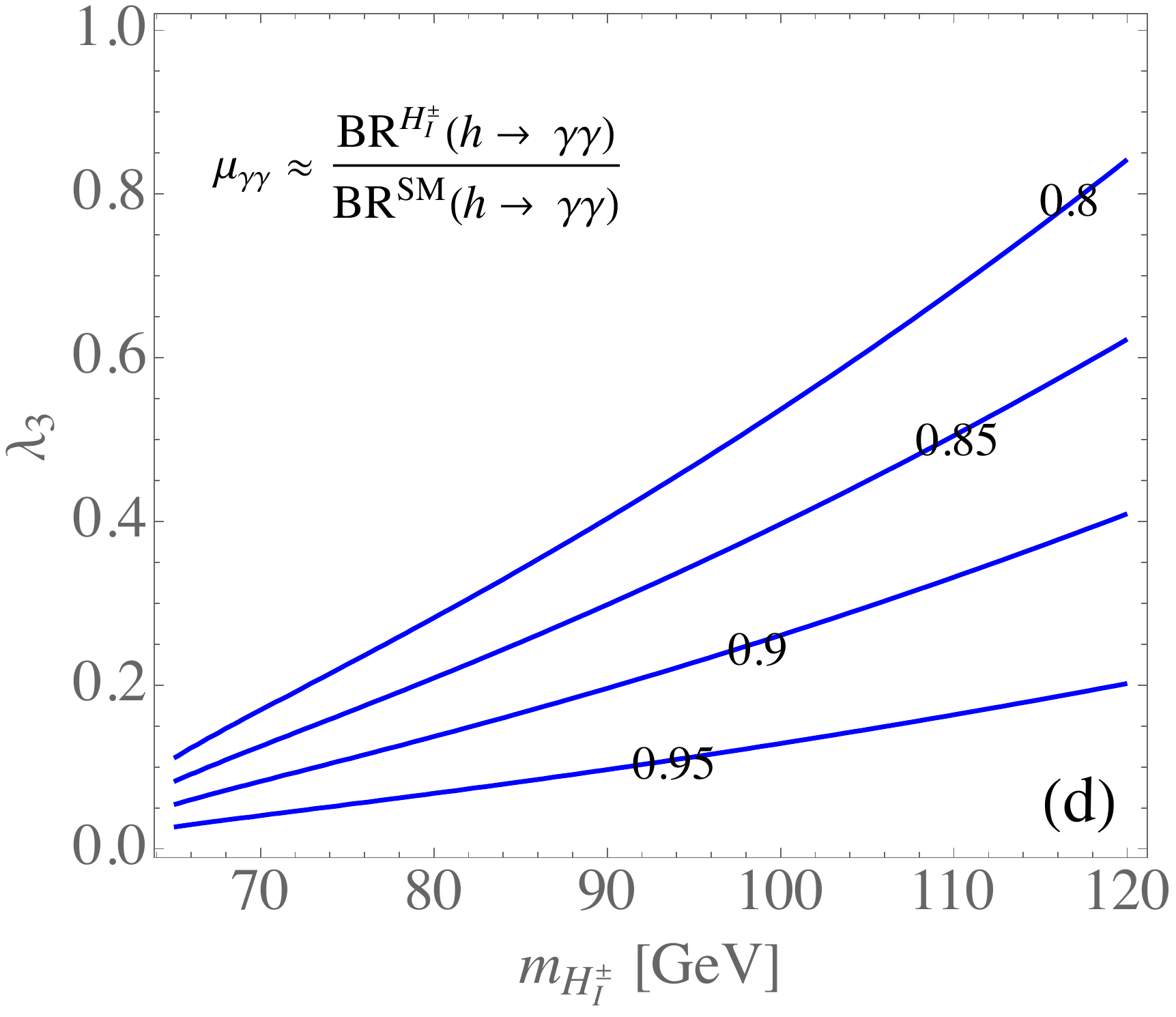}
 \caption{ Plots (a)-(c) show  the constraints from the oblique $T$-parameter, and plot (d) is the constraint from $\mu_{\gamma\gamma}$.  }
\label{fig:bounds}
\end{center}
\end{figure}

The mass regions in Eqs.~(\ref{eq:Qmass_region}) and (\ref{eq:smass_region}) are only used to set the boundaries of the scanned parameters.  Their mass differences are dictated by the oblique $T$ parameters, as shown in Eqs.~(\ref{eq:Q_T}) and (\ref{eq:S_T}). To understand the $T$-parameter constraints, we perform the parameter scan based on the chosen parameter regions. The  correlation of $m_{B_2}-m_{T}$ and $m_{B_2} -m_{B_1}$  is shown in Fig.~\ref{fig:bounds}(a), where $5\times 10^{6}$ random sampling points are used. The allowed region for $m_{B_2}-m_T$ is limited within  200 GeV and  that for $m_{B_2}-m_{B_1}$ is wider; that is, the mass splitting in the same representation is strictly constrained.  Therefore, it is appropriate if we take  $m_{B_2} -m_T =100$ GeV  for the top decay analysis. Similarly, the $T$-parameter constraint on $m_{A_I}-m_{H_I}$ and $m_{H^\pm_I-H_I}$ is shown in Fig.~\ref{fig:bounds}(b), where the mass ordering $m_{H^\pm} > m_{A_I} > m_{H_I}$ is applied. Because the chosen ranges of  $m_{H_I, A_I, H^\pm_I}$ in Eq.~(\ref{eq:smass_region}) are not broad, the constraint is not significant.  As $m_{B_2}-m_{T}$ is related to the parameter $y_{B'}$ or $s_\theta$,  the correlation between  $y_{B'}$ and $s_\theta$ under the $T$-parameter constraint is shown in Fig.~\ref{fig:bounds}(c). When  $m_{B_2}-m_T$ is limited,  the allowed $y_{B'}$ parameter is bounded by $|y_B'|\lesssim 1.5$, where the maximum $s_\theta$ can still reach $|s_\theta|_{\rm max} \sim 1/\sqrt{2}$. Using Eq.~(\ref{eq:mu2gamma_ICH}),  the strength signal for the Higgs to diphoton as a function of $m_{H^\pm_I}$ and $\lambda_3$ is shown in Fig.~\ref{fig:bounds}(d). If the uncertainty of the observed strength signal is $10\%$ of the SM result, then the $\lambda_3$ value is limited to be approximately $0.2$ when $m_{H^\pm}=90$ GeV is used.

\subsection{ BRs for $t\to q (h ,\gamma)$ and $t\to q Z$}

After analyzing the constraints of the parameters, in this subsection, we numerically calculate and discuss the $S_I$- and $H^\pm_I$-mediated contributions to the $t\to q (h,\gamma)$ and $t\to qZ$ decays.  As stated before, many parameters are involved in the processes. We can use the parameter scan  to comprehend the influence of various parameters. Before scanning the parameters, we need to examine whether  the $t\to q V$ and $t\to q h$  decays can be simultaneously enhanced  to the current experimental upper limits. 

\begin{figure}[phtb]
\begin{center}
\includegraphics[scale=0.4]{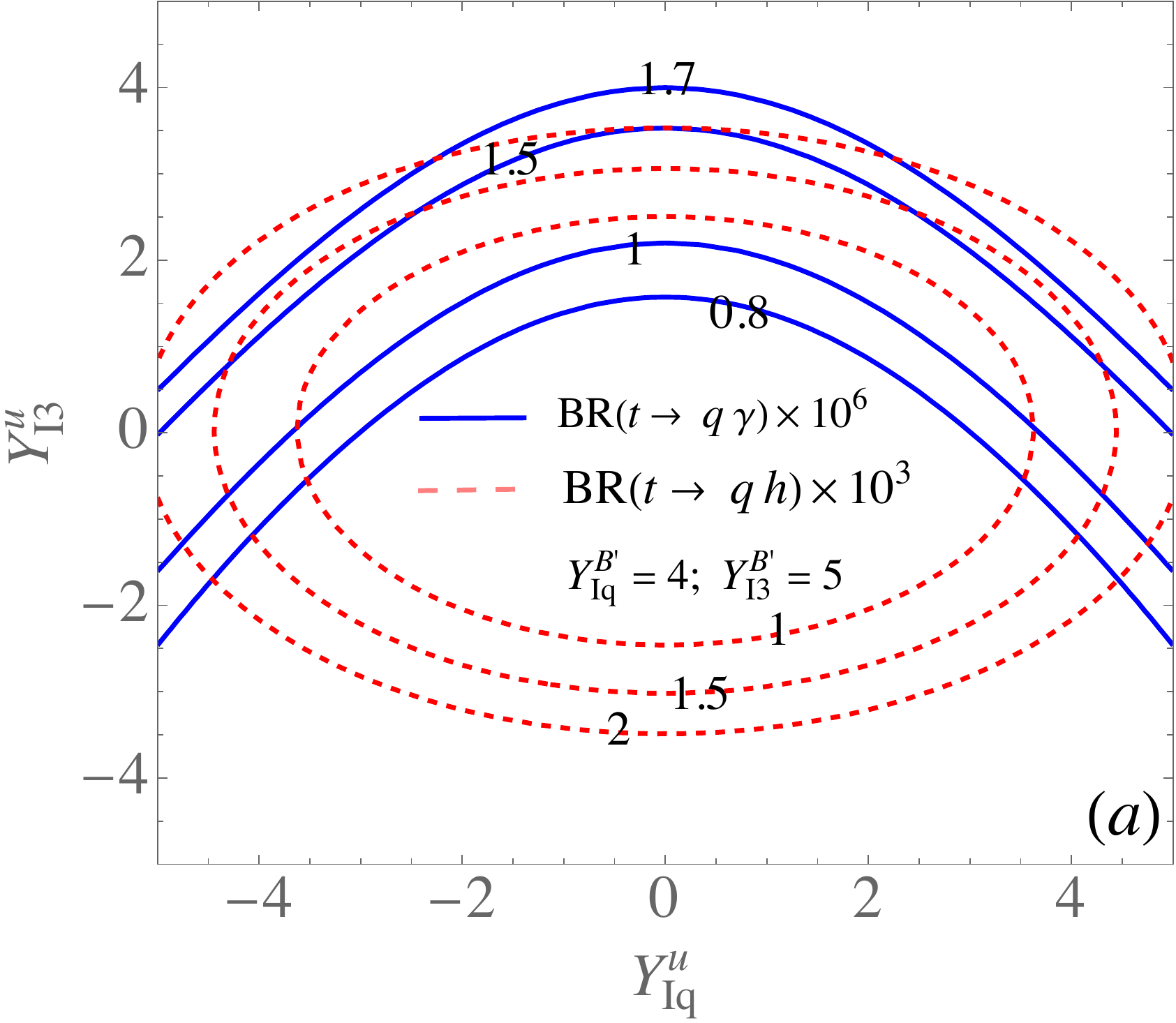}
\includegraphics[scale=0.4]{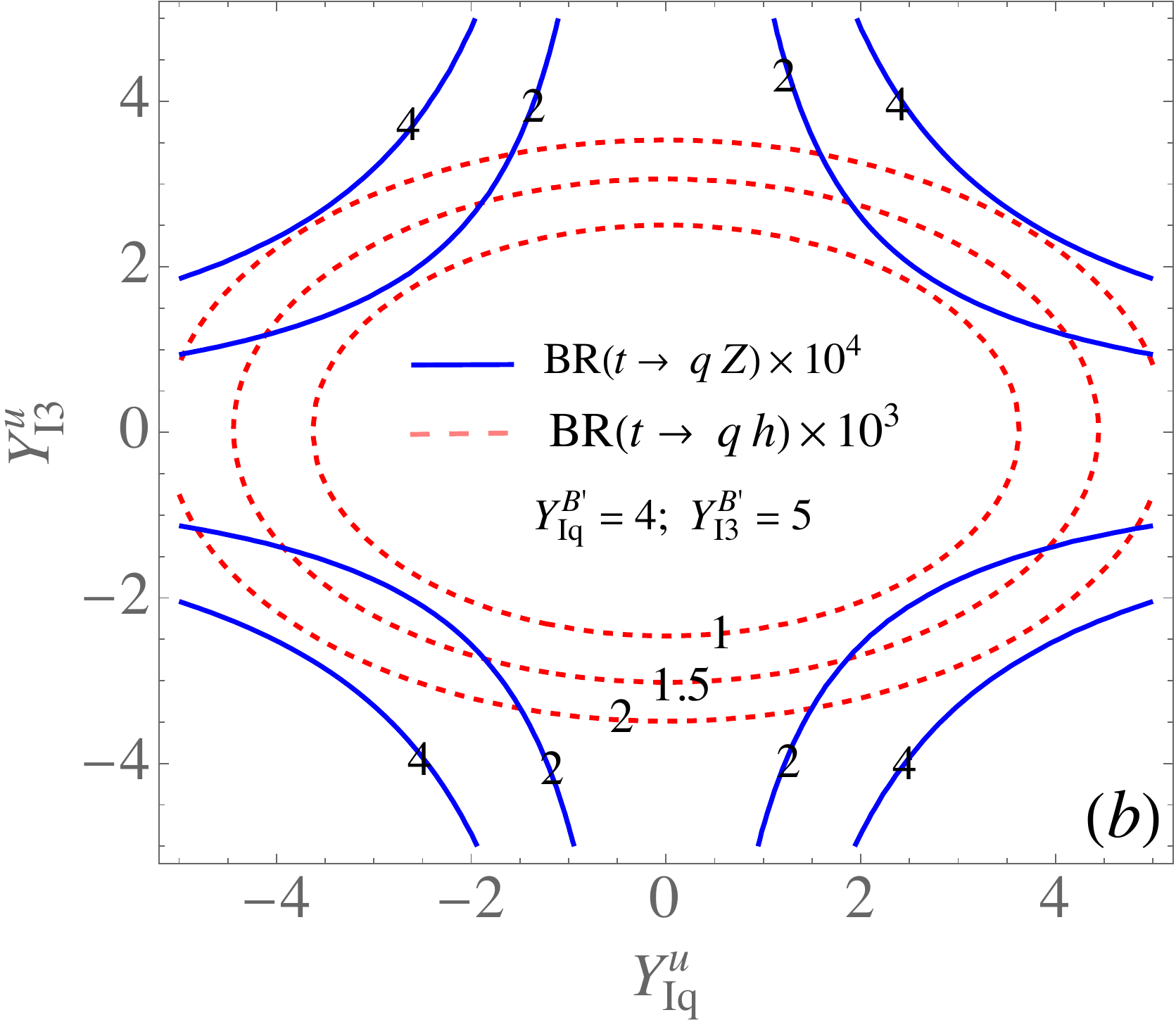}
 \caption{Contours for (a) $BR(t\to q \gamma)$ and (b) $BR(t\to qZ)$  as a function of  $Y^{u}_{Iq}$ and $Y^{u}_{I 3}$, where the dashed lines are $BR(t\to q h)$ and $Y^{B'}_{Iq,I3}=(4,5)$ are used. The other taken parameter values can be found in the text.}
  \label{fig:qV-qh}
\end{center}
\end{figure}

\begin{figure}[phtb]
\begin{center}
\includegraphics[scale=0.35]{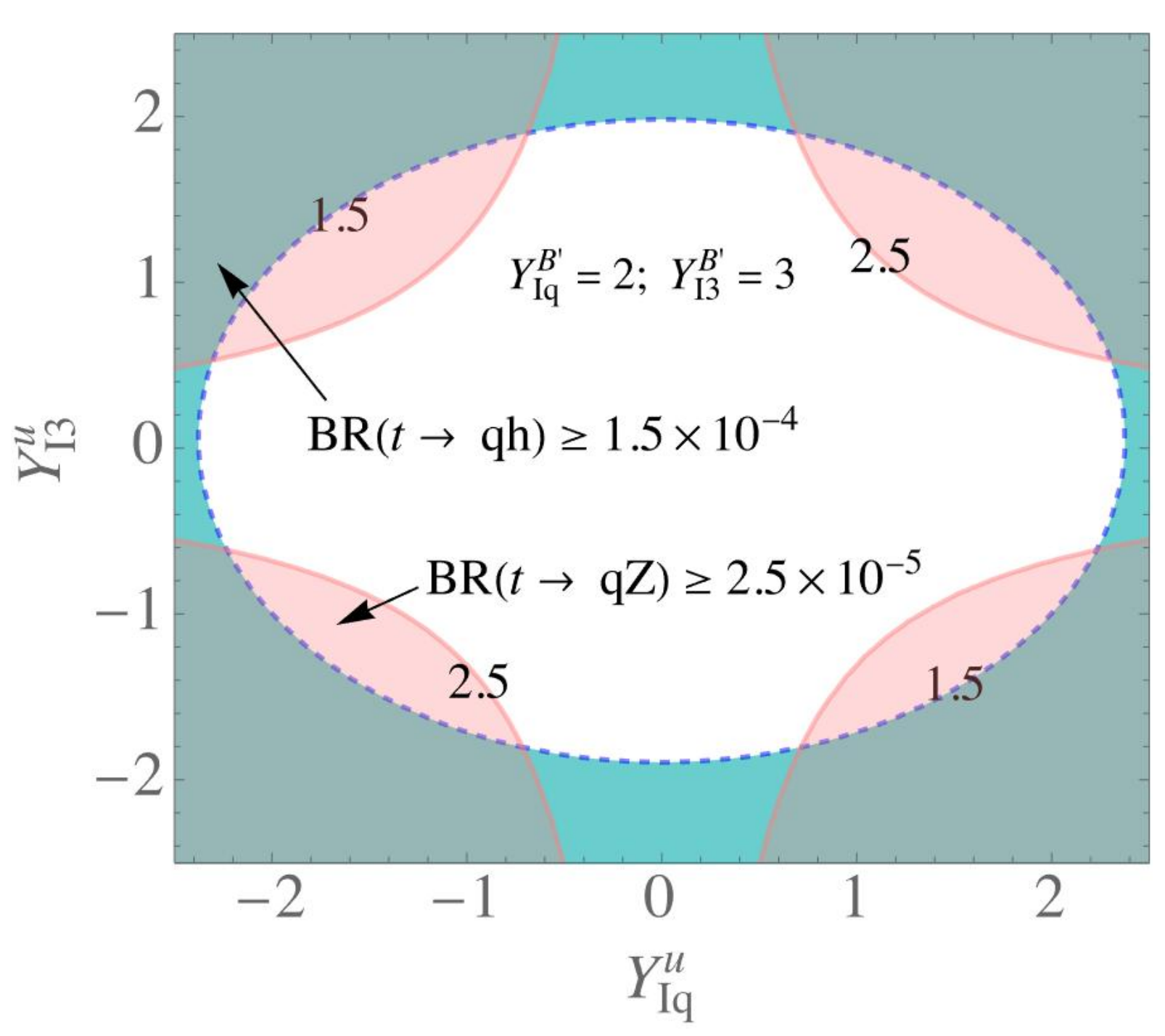}
 \caption{Regions for $BR(t\to q h)\geq 1.5\times 10^{-4}$ and $BR(t\to qZ)\geq 1.5 \times 10^{-5}$ as a function for $Y^{u}_{Iq}$ and $Y^{u}_{I3}$, where $Y^{B'}_{Iq}=2$, and $Y^{B'}_{I3}=3$ are used.}
  \label{fig:qh-qZ}
\end{center}
\end{figure}

For the purpose of illustration, we use the formulas given in Eqs.~(\ref{eq:BRtqga}) and (\ref{eq:BRtqZ}) and show the contours for $BR(t\to q \gamma)$ (in units of $10^{-6}$)  and $BR(t\to q Z)$ (in units of $10^{-4}$) in Fig.~\ref{fig:qV-qh}(a) and (b), respectively, where  the dashed lines denote  $BR(t\to q h)$. The parameter values are taken as  $m_{H_I, A_I, H^\pm_I}=(65, 70, 90)$ GeV, $m_{B_1,B_2,T}=(1, 1.5,1.4)$ TeV, $\lambda_3=0.2$, $y_{B'}=1.0$, and $Y^{B'}_{Iq,I3}=(4,5)$. As shown in  the plots,  the BRs for $t\to q (h, \gamma, Z)$ with the chosen parameter values can reach the levels of $(10^{-3}, 10^{-6}, 10^{-4})$, where with the exception of $t\to q\gamma$, $t\to q (h, Z)$ can reach the current upper bounds that are shown in Eq.~(\ref{eq:exp_upper}). Based on the results, although the used Yukawa couplings are lower than the upper limit from the perturbative unitarity, $BR(t\to q \gamma)$ of $O(10^{-6})$ inevitably  has to reply  on the large Yukawa couplings. To illustrate the situation with small  Yukawa couplings, we fix $Y^{B'}_{Iq,I3}=(2,3)$  and show the regions for $BR(t\to q h)\geq 1.5 \times 10^{-4}$ and $BR(t\to q Z)\geq 1.5\times 10^{-5}$ as a function of $Y^u_{Iq, I3}$ in Fig.~\ref{fig:qh-qZ}. In these chosen regions,  $BR(t\to q \gamma)$ is far below $10^{-6}$. 

In the following, we discuss the influence of various parameters on the rare top decays in detail. To reduce the number of scanned parameters,  the parameters that are  insensitive to the studying phenomena are fixed  as follows
 \begin{equation}
 m_{H_I, A_I, H^\pm_I}=(65, 70, 90)~\text{GeV}\,,~ \lambda_3=0.2\,,
 \end{equation}
where the  values are chosen to satisfy the constraints obtained earlier.  Thus, the  involving parameters are the $Z_2$-odd quark masses $m_{B_1, B_2, T}$ and Yukawa couplings $y_{B'}$, $Y^{B',u}_{I1,I3}$, and their  scanning regions  are chosen as
  \begin{align}
  m_{B_1} &  \in [1000, 1300]~{\rm GeV}\,,~m_{T}\in [1000, 2000]~{\rm GeV}\,,~m_{B_2}=m_T + 100\,, \nonumber \\
  y_{B'} &\in (-1.5,1.5)\,, ~Y^{B'}_{Iq,I3}   \in (-3, 3)\,,~Y^{u}_{Iq,I3}\in (-3, 3)\,. \label{eq:para_range}
   \end{align}
According to the ATLAS results of $BR(t\to u g) \lesssim 0.61\times 10^{-4}$~\cite{ATLAS:2021amo}, we can indirectly bound the $t\to q \gamma$ to be $BR(t\to q\gamma) \lesssim 3.2 \times 10^{-6}$  using $BR(t\to q \gamma) \sim BR(t\to q\gamma)\alpha/(C_F \alpha_s)$ in the model, where the upper limit is smaller than the current experimental  upper limit.  Therefore, it is sufficient to use the small Yukawa couplings for the scan. In addition, in order to show the resulting BRs that can reach the sensitivities at the HL LHC, we require the obtained BRs for the rare top decays to be
 \begin{align}
 10^{-5} &< BR(t\to q h) < 10^{-3}\,, \nonumber \\
  2 \times 10^{-5} &< BR(t\to q Z) < 5 \times 10^{-4}\,. \label{eq:BR_range}
 \end{align}
However, the radiative top decay is required to be $BR(t\to q \gamma) > 0.3 \times 10^{-6}$. Since we have taken small Yukawa couplings, $BR(t\to q \gamma)$ cannot reach the current upper limit, i.e., $O(10^{-4})$. Thus, it is not necessary to set the upper value for $BR(t\to q \gamma)$.  

\begin{figure}[phtb]
\begin{center}
\includegraphics[scale=0.45]{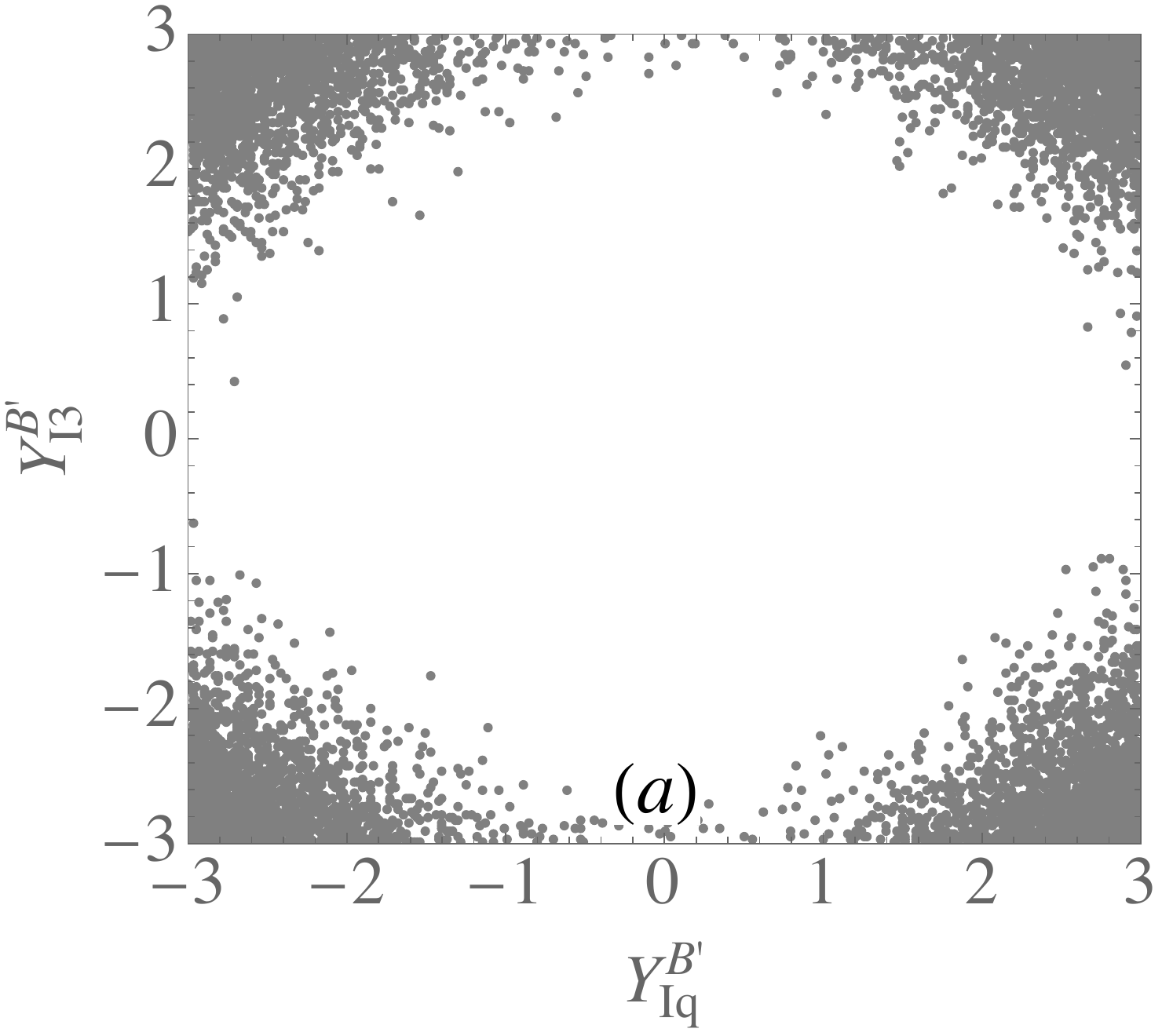}
\includegraphics[scale=0.45]{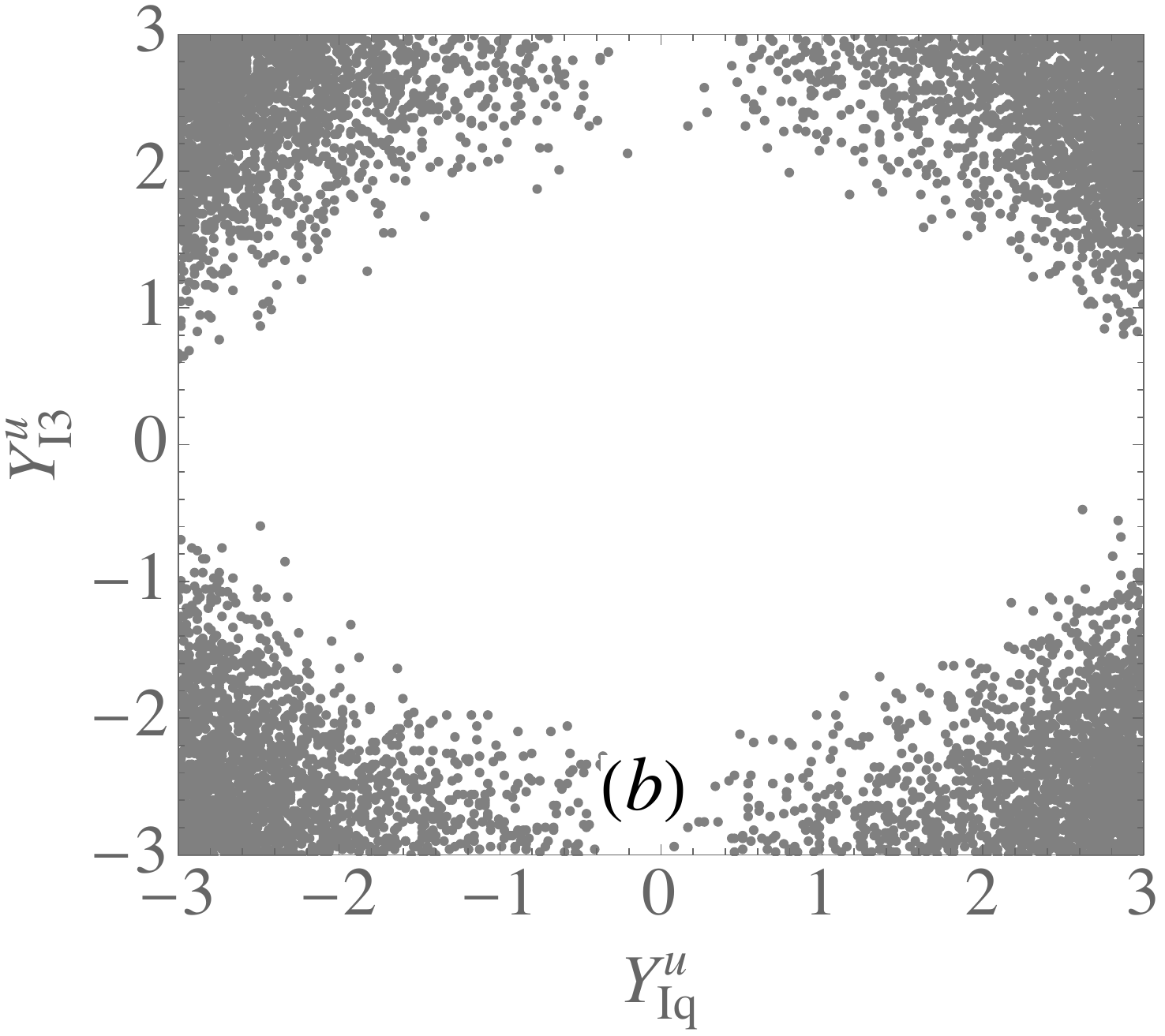}
\includegraphics[scale=0.45]{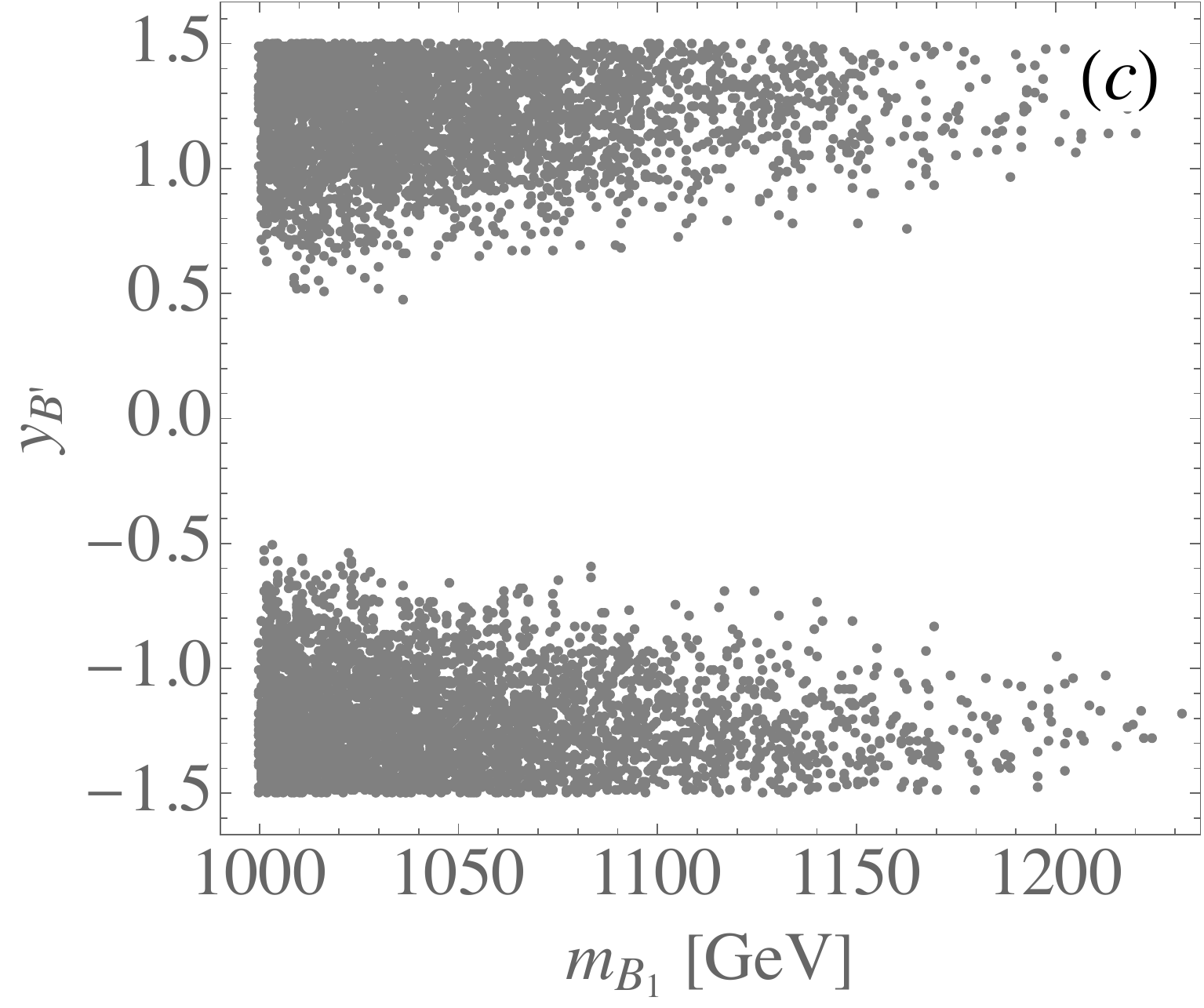}
\includegraphics[scale=0.45]{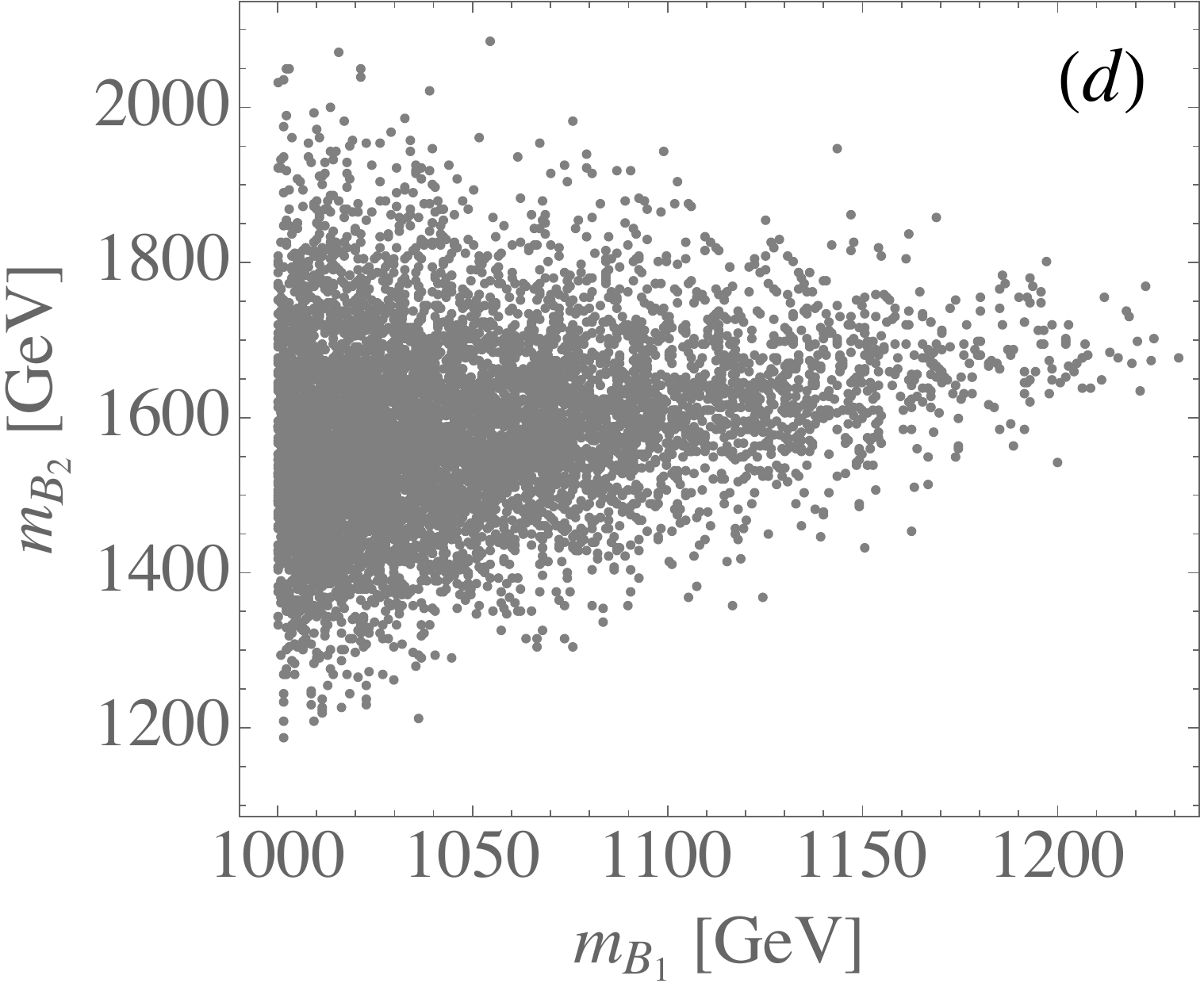}

 \caption{ Correlations of Yukawa couplings and $m_{B_{1,2}}$ that fit the taken ranges shown in Eqs.~(\ref{eq:para_range}) and (\ref{eq:BR_range}). }
\label{fig:YBp_Yu}
\end{center}
\end{figure}

We show the scatter plots for the Yukawa couplings of $Y^{B'}_{Iq, I3}$ and $Y^u_{Iq,I3}$, which fit the  ranges given in Eqs.~(\ref{eq:para_range}) and (\ref{eq:BR_range}), in Fig.~\ref{fig:YBp_Yu}(a) and (b). Hence,  $|Y^{B',u}_{Iq, I3}|<1$ are disfavored.  The correlation between $y_{B'}$ and $m_{B_1}$ is shown in Fig.~\ref{fig:YBp_Yu}(c). From the numerical results, $|y_{B'}|\lesssim 0.5$  is excluded. In addition, the correlation between $m_{B_1}$ and $m_{B_2}$ can be found  in Fig.~\ref{fig:YBp_Yu}(d). 

\begin{figure}[phtb]
\begin{center}
\includegraphics[scale=0.4]{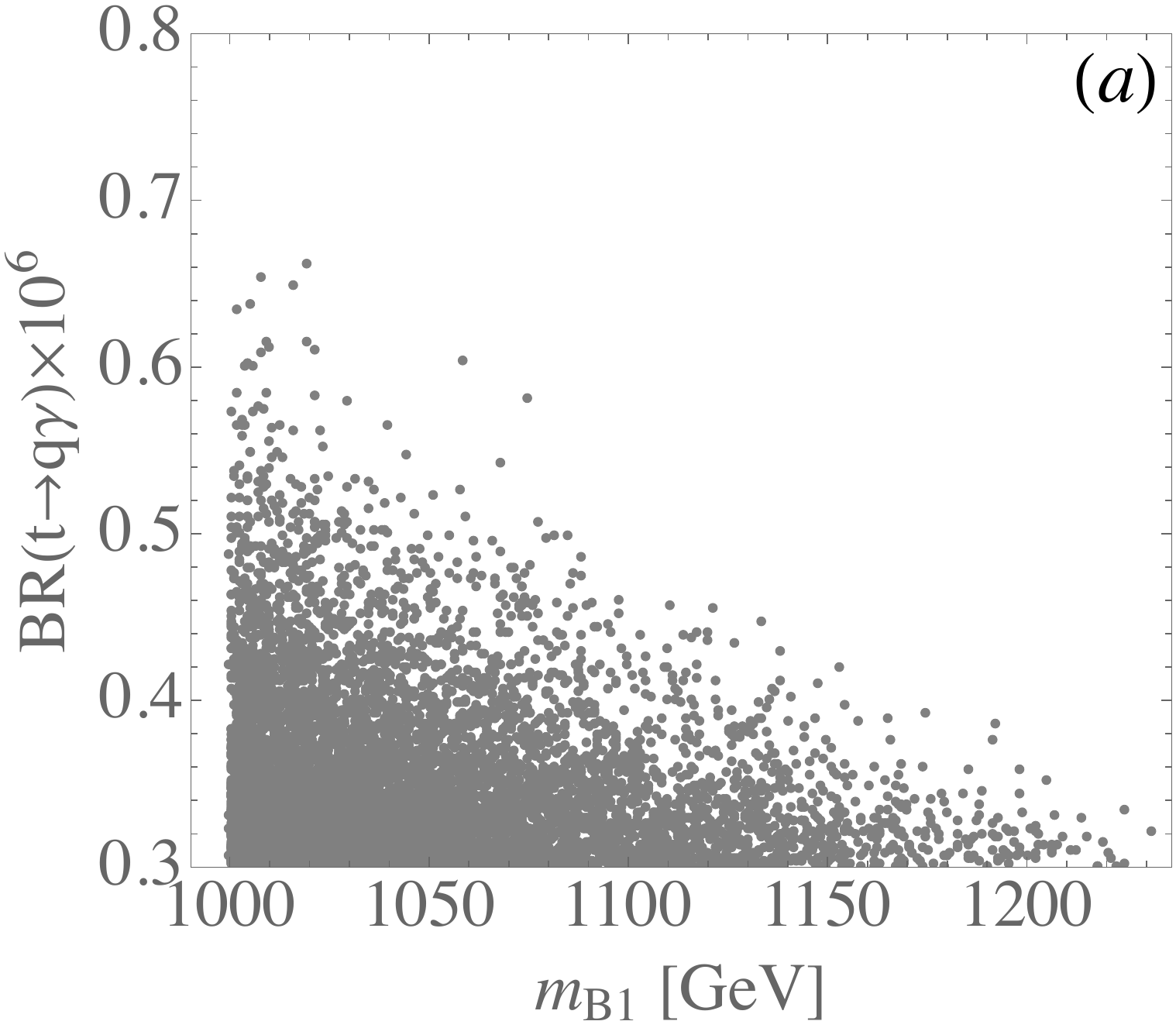}
\includegraphics[scale=0.4]{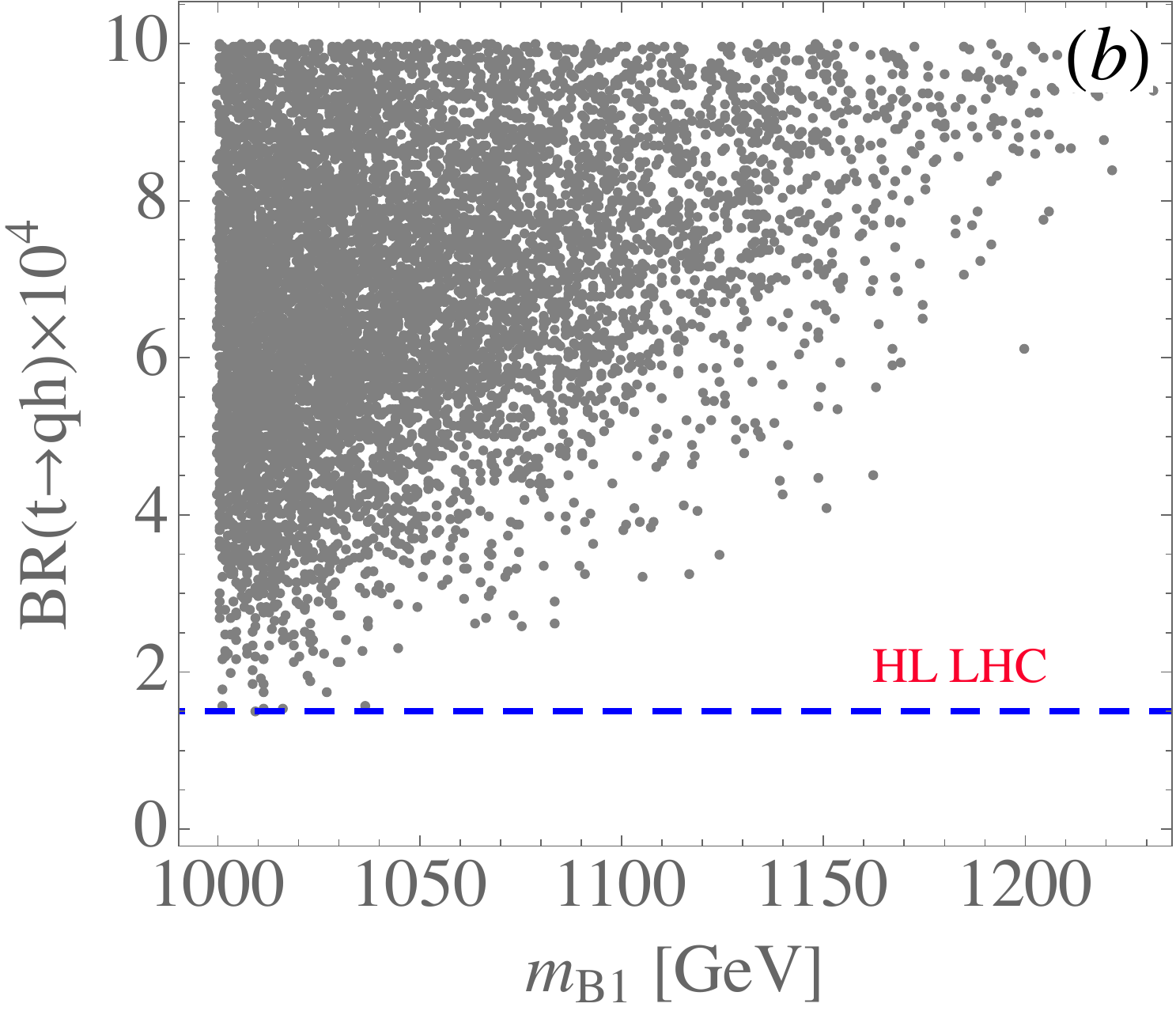}
\includegraphics[scale=0.4]{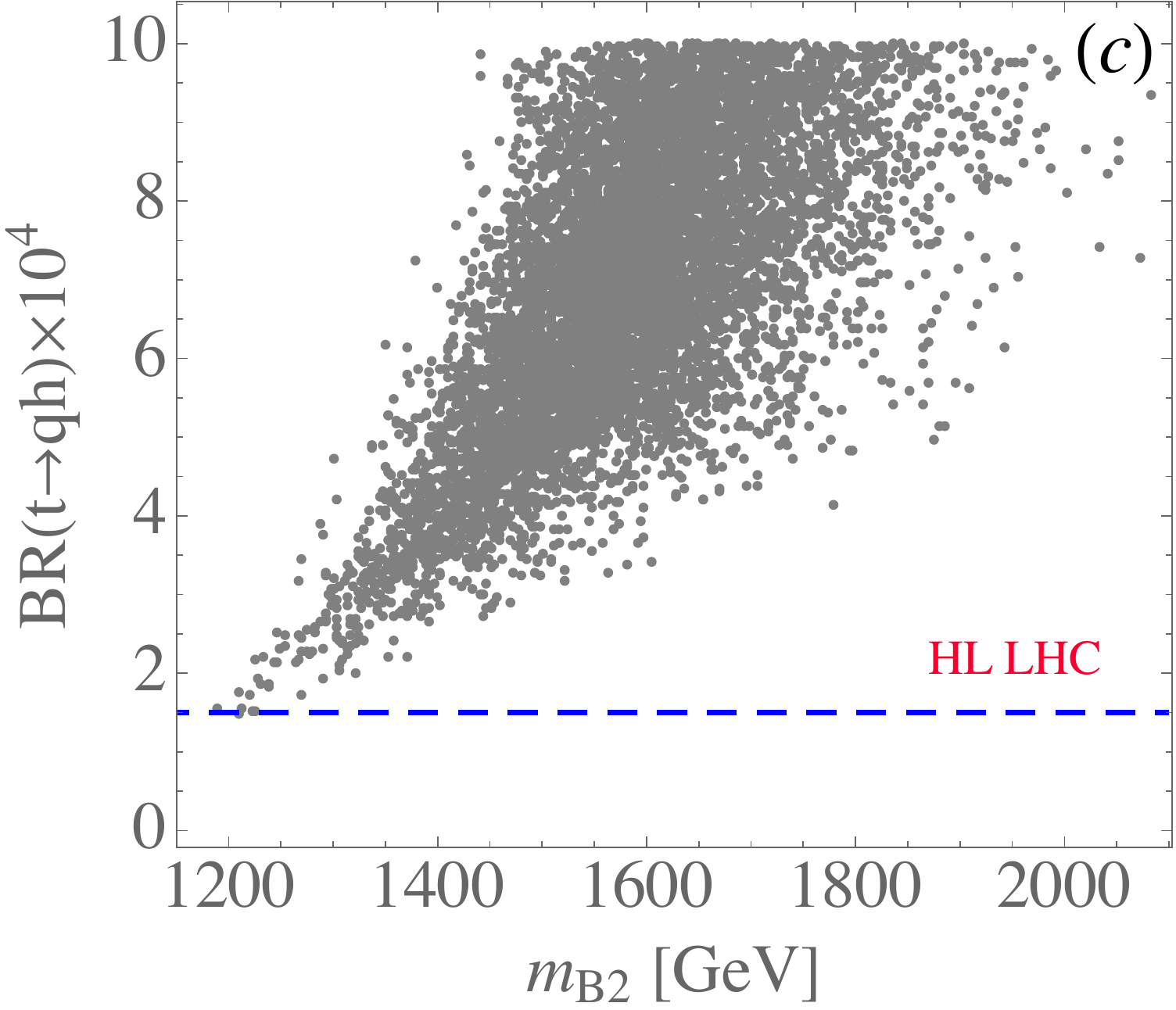}
\includegraphics[scale=0.4]{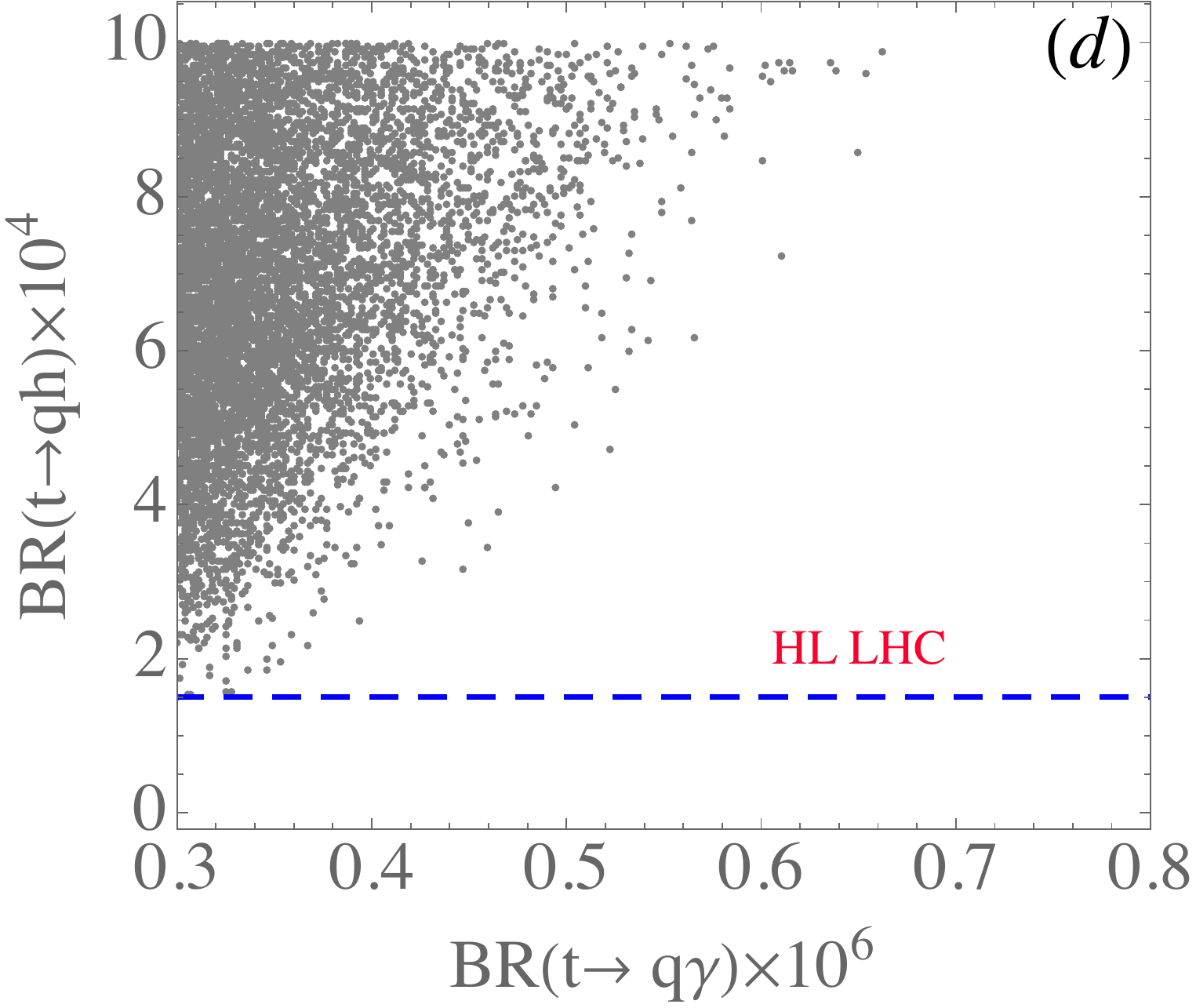}
 \caption{  (a)-(b) $BR(t\to q \gamma, q h)$ as a function of $m_{B_1}$; (c) $BR(t\to q h)$ as a function of $m_{B_2}$, and (d) correlation between $BR(t\to q \gamma)$ and $BR(t\to q h)$. }
\label{fig:h-ga}
\end{center}
\end{figure}

After knowing the constrained parameter ranges, in the following, we discuss the implications on the $t\to q (\gamma,h, Z)$ decays. Because $t\to q\gamma$ and $t\to q h$ have similar chirality structures in decay amplitudes and arise from the similar Feynman diagrams, we analyze their numerical results together.  When neglecting the small contributions from Fig.~\ref{fig:ttoqh}(c) and (d),  as stated earlier, $t\to q \gamma, q h$ decays can arise  from Fig.~\ref{fig:ttoqh}(a) and (b). Compared to the $H^\pm \gamma$ coupling, the $B_i B_i \gamma$ coupling has a suppression factor from the electric charge of $B_i$; thus, the $t\to q \gamma$ decay indeed is dominated by Fig.~\ref{fig:ttoqh}(a). By contrast, because the $H^\pm_I h$ coupling is $\lambda_3$ and is limited by the $h\to \gamma\gamma$ measurement, the dominant contribution to $t\to q h$ is from Fig.~\ref{fig:ttoqh}(b).  Hence, the resulting $BR(t\to q \gamma)$ and $BR(t\to q h)$ as a function of $m_{B_1}$ are shown in Fig.~\ref{fig:h-ga}(a) and (b).  In the chosen Yukawa couplings, the BR for $t\to q \gamma$  can only maximally reach of the $O(10^{-7})$, which is consistent with the indirect bound from the $t\to q g$ measurement.  The dependence of $B(t\to q h)$ with respect to $m_{B_2}$ is shown in Fig.~\ref{fig:h-ga}(c), where  $BR(t\to q h)$ increases as $m_{B_2}$ increases.  The reason for the behavior can be understood as the relaxed cancellation between  $B_1$ and $B_2$ when the mass of $B_2$ increases; that is, the $B_1$ becomes the dominant effect if $m_{B_2} \gg m_{B_1}$.  In Fig.~\ref{fig:h-ga}(d), we show the scatter plot for the correlation between $BR(t\to q \gamma)$ and $BR(t\to qh)$. The dashed lines in Fig.~\ref{fig:h-ga} denote the sensitivity of the HL LHC.

\begin{figure}[phtb]
\begin{center}
\includegraphics[scale=0.5]{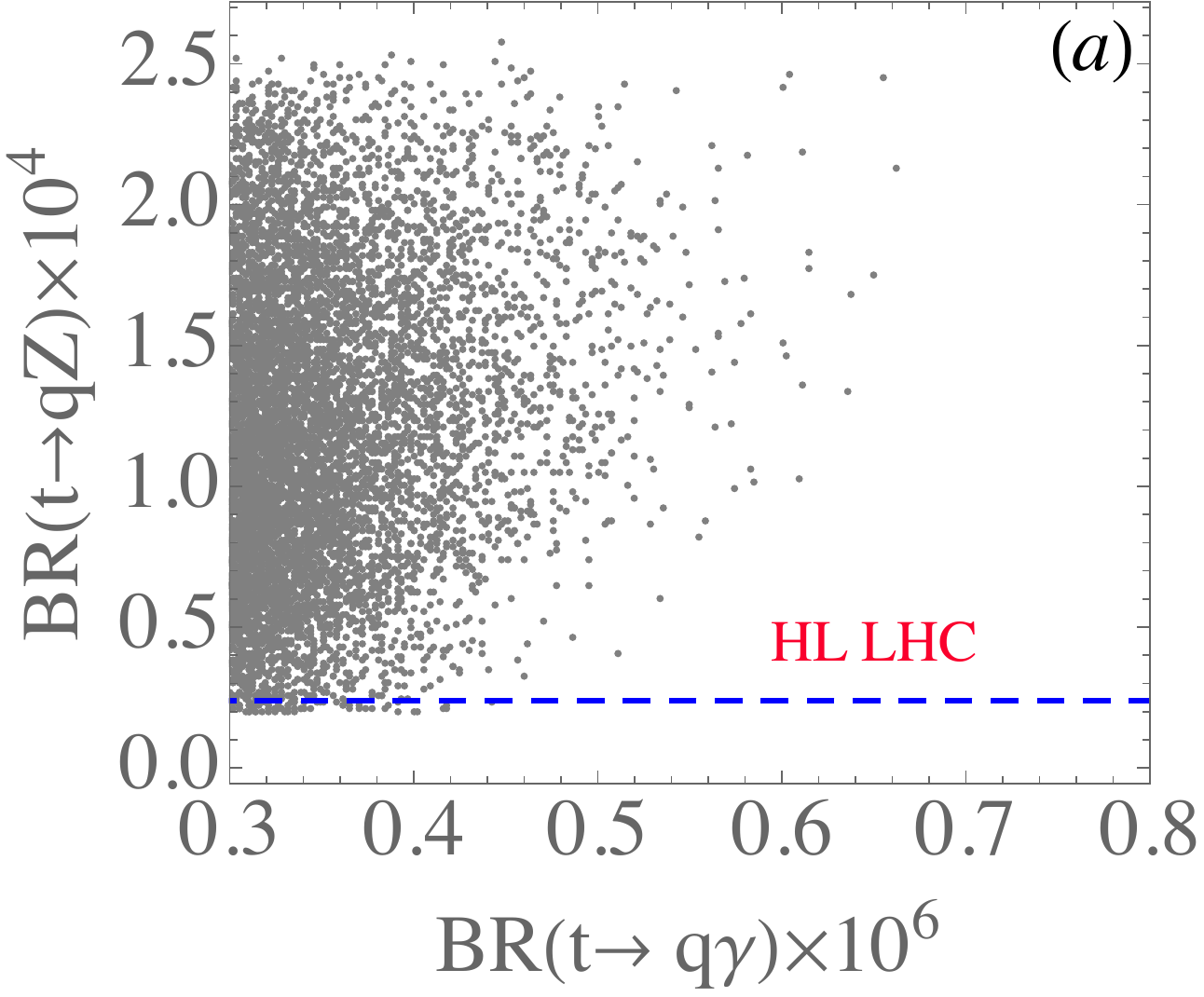}
\includegraphics[scale=0.5]{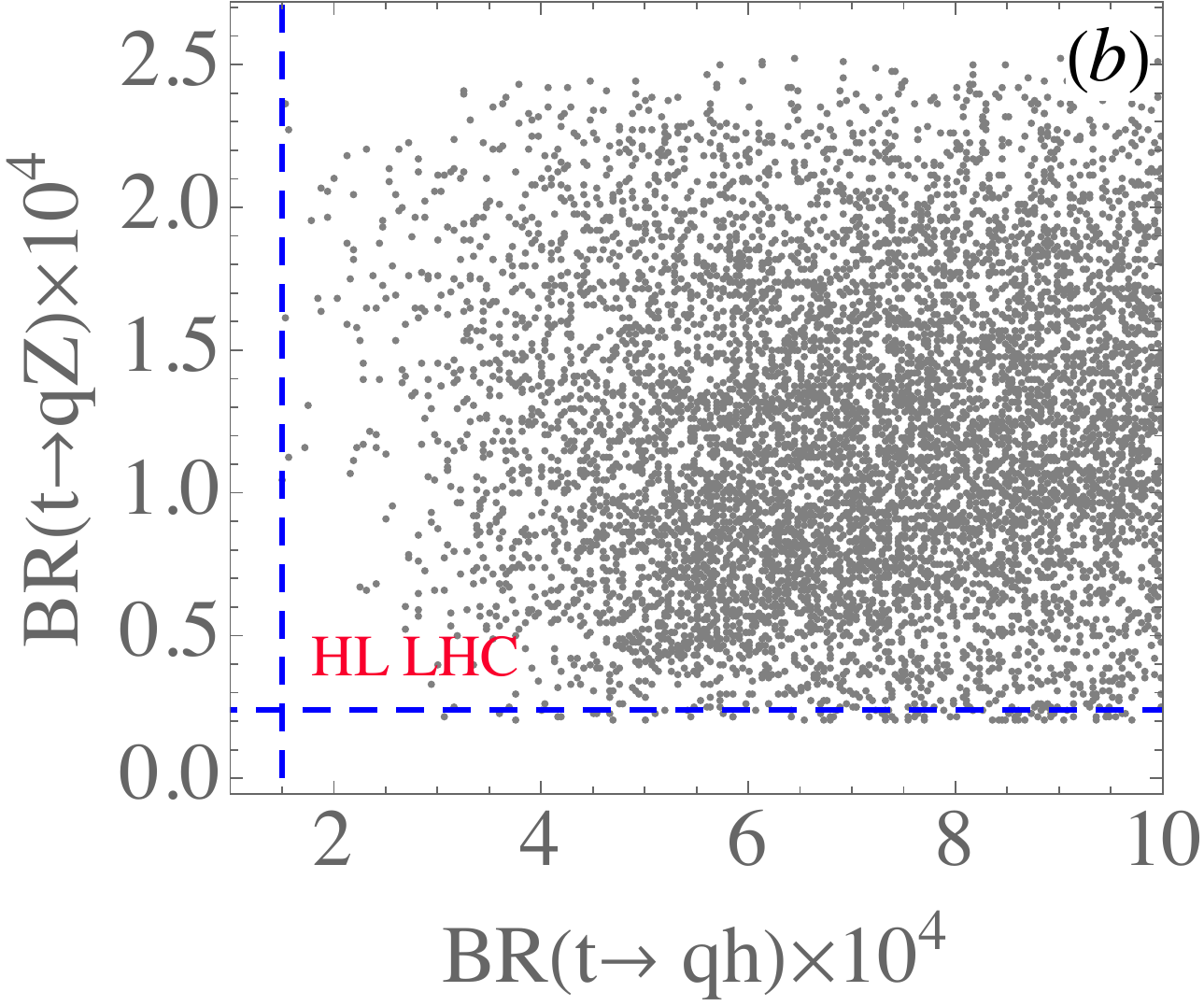}
 \caption{ Correlation of $BR(t\to q Z)$ with  (a) $BR(t\to q \gamma)$ and (b) $BR(t\to qh)$.  }
\label{fig:BrqZ}
\end{center}
\end{figure}

In addition to the dipole operators, unlike the $t\to q\gamma$ decay, $t\to q Z$ also involves the vectorial types of operators  in the  decay amplitude, which are not associated with the chirality flip in the quark currents, so  they are the dominant effect. To observe the influence of the new physics effects on $t\to q Z$, we show the scatter plots for the correlations of $BR(t\to q Z)$ with $BR(t\to q \gamma)$ and with $BR(t\to q h)$ in Fig.~\ref{fig:BrqZ}(a) and (b), respectively, where the dashed lines are the sensitivity that the  HL LHC is planning to reach. With the exception of the $t\to q \gamma$ decay, the BRs for the $t\to q h$ and $t\to q Z$ decays  in the model can reach the level of  $O(10^{-4})$. 

\section{Summary} \label{sec:summary}

In this study, we investigated the potential effects that can  enhance the top-FCNC processes, where the processes  are highly suppressed in the SM. If we assume that $t\to q h$ and $t\to q Z$ are induced via quantum loop diagrams, the intermediate states in the loop may have different properties from the SM particles. Inspired from the mechanism of the scotogenic  neutrino mass,  we consider that the new particles carry an extra $Z_2$-odd parity, whereas the SM particles are $Z_2$-even. 

To retain the basic element in the radiative neutrino mass~\cite{Ma:2006km},  in addition to the $Z_2$ discrete symmetry, we can extend the SM by including one inert Higgs doublet, one vector-like $Z_2$-odd doublet quark, and one vector-like $Z_2$-odd singlet quark.

The potential constraints from the experimental observations are taken into account, such as the oblique parameters, Higgs to diphoton decay, and DM direct detection. Although  the $b\to s\gamma$ decay and $\Delta B=2$ process can be induced in the model, their effects can be small. 

According to the recent ATLAS's  measurement that shows $BR(t\to q g)\lesssim O(10^{-4})$, we found that the indirect bound on $t\to q \gamma$ in the model is $BR(t\to q \gamma) \lesssim 3.2 \times 10^{-6}$, which is lower than the expected sensitivity at the HL LHC. With the exception of $t\to q \gamma$, the branching ratios for the loop-induced $t\to q h$ and $t\to q Z$ decays in the model can be of the order of $10^{-4}$ and can be tested at the HL LHC.

\section*{Acknowledgments}

This work was supported in part by the Ministry of Science and Technology, Taiwan under the Grant No.~MOST-110-2112-M-006-010-MY2 (C.~H.~Chen).
The work was supported in part by the Fundamental Research Funds for the Central Universities (T.~N.).

\end{document}